\documentclass{iopjournal}
\usepackage{amsmath}
\usepackage{amssymb}

\begin{document}
\articletype{Paper} 
\title{Six Open Questions in Machine-Learned Interatomic Potential Foundation Models}

\author{
Isabel Creed$^{1, 9, *}$\orcid{0009-0001-3930-7873},
Tim Rein$^{2, 1, 9, *}$\orcid{0009-0004-9681-0055},
Ingvars Vitenburgs$^{1, 9, *}$\orcid{0000-0002-5816-9998},
Wojciech G. Stark$^{1, 9, *}$\orcid{0000-0001-6279-2638},
Viktor Ellingsson$^{1, 9}$\orcid{0009-0005-3858-5491},
Ahmed Y. Ismail$^{3, 9}$\orcid{0009-0006-6782-6319},
Guangyu Liu$^{2, 9}$,
Yuchen Lou$^{1, 9}$,
Bradley A. A. Martin$^{3, 9}$\orcid{0000-0003-1583-254X},
Cyprien Bone$^{3, 9}$\orcid{0009-0002-7801-5272},
Matthew A. H. Walker$^{3, 9}$\orcid{0000-0001-8746-8754},
Mueen Taj$^{3, 9}$,
Shirui Wang$^{1, 9}$\orcid{0009-0008-6657-9971},
Kelvin Wong$^{4, 9}$\orcid{0009-0003-4036-2727},
Ruiqi Wu$^{1, 9}$\orcid{0009-0006-3736-3958},
Prakriti Kayastha$^{3, 9}$\orcid{0000-0002-4852-6445},
Bingqing Cheng$^{7}$\orcid{0000-0002-3584-9632},
Aditi Krishnapriyan$^{7, 10, 11}$\orcid{0000-0003-3472-6080},
Michele Ceriotti$^{8}$\orcid{0000-0003-2571-2832},
Marcel F. Langer$^{8}$\orcid{0000-0002-1270-3016},
Jarvist Moore Frost$^{1, 9, *}$\orcid{0000-0003-1938-4430},
Alex M. Ganose$^{1, 9, *}$\orcid{0000-0002-4486-3321},
Venkat Kapil$^{5, 6, 9, *}$\orcid{0000-0003-0324-2198},
Keith T. Butler$^{3, 9, *}$\orcid{0000-0001-5432-5597}
}

\affil{$^1$
Department of Chemistry, 
Imperial College London, 
South Kensington Campus, 
London, SW7 2AZ, UK.}\\
\affil{$^2$
Department of Physics, 
Imperial College London, 
South Kensington Campus, 
London, SW7 2AZ, UK.}\\
\affil{$^3$
Department of Chemistry, 
University College London, 
Kathleen Lonsdale Building, 
Gower Pl, 
London, WC1E 6BS, UK.}\\
\affil{$^4$
Department of Chemical Engineering,
University College London,
Roberts Building,
Torrington Pl,
London, WC1E 7JE, UK.}\\
\affil{$^5$
Department of Physics and Astronomy,
University College London,
7--19 Gordon St,
London, WC1H 0AH, UK.}\\
\affil{$^6$
London Centre for Nanotechnology,
University College London,
9 Gordon St,
London, WC1H 0AH, UK.}\\
\affil{$^7$
College of Chemistry, 
University of California Berkeley, 
CA 94720-146, USA.}\\
\affil{$^8$
Laboratory of Computational Science and Modeling,
Institute of Materials,
École Polytechnique Fédérale de Lausanne,
1015 Lausanne, Switzerland.}\\
\affil{$^9$
The Thomas Young Centre,
London WC1E 6N, UK.}\\
\affil{$^{10}$
Department of Electrical Engineering and Computer Sciences,
University of California Berkeley, 
CA 94720-146, USA.}\\
\affil{$^{11}$
Applied Mathematics and Computational Research Division,
Lawrence Berkeley National Laboratory.}

\email{k.t.butler@ucl.ac.uk}

\begin{abstract}
Machine-learned interatomic potentials (MLIPs) have had a profound impact on molecular modelling in recent years, promising to resolve the long-standing tension between the scale and accuracy of simulations. There has been a proliferation of new models and designs, and recently the paradigm of ``foundational'' MLIPs has become prevalent. Broadly speaking, foundation models are trained on large diverse datasets and promise to work well for new systems with minimal updates required. However, in such a new and fast moving field, there are many unanswered questions. In this article, we set out to articulate and explore what we see as the most important among these questions. We start by developing a working definition for foundational MLIPs and use this definition to frame the subsequent open questions. Despite the rapid progress in the field of MLIP models, we believe that these are fundamental questions which will continue to define cutting edge research in MLIPs in the years to come. 
\end{abstract}

\section*{Introduction}

Machine learning (ML) has rapidly become a central tool in atomistic modelling, reshaping how molecular and materials systems are described and explored. 
The pace of methodological development has accelerated dramatically, making it increasingly difficult even for specialists to maintain a coherent overview of the field.
Against this backdrop, it is timely to take stock of areas where ML has moved beyond proof-of-concept studies to become a practical component of everyday atomistic workflows. 
Among the many strands of research within ML-driven atomistic modelling, one development stands out for both its speed of maturation and its breadth of impact: universal machine-learned interatomic potentials. 
In only a few years, these models have evolved from an emerging research direction into a widely adopted technology, enabling new scales of simulation and opening up fresh opportunities in science and innovation.



Interatomic potentials have a venerable history in the field of molecular and materials simulation. An interatomic potential describes the potential energy of a set of interacting atoms. 
This can be combined with statistical sampling techniques such as Monte Carlo methods~\cite{Metropolis1953,Hastings1970} to predict equilibrium thermodynamic properties of molecular and periodic systems.
%
%
Taking the negative derivative of this potential with respect to the atomic positions provides atomic forces (the ``force field''). 
This force field can then be used with a numeric integrator to solve Newton's equation of motion, simulating the classical molecular dynamics trajectories and associated dynamical and thermodynamic observables~\cite{Rahman1964,Verlet1967}.
%
The same force field can be used with imaginary-time path-integral molecular dynamics to describe nuclear quantum effects~\cite{Chandler1981,Parrinello1981,Ceperley1995}.
%
These interatomic potentials have driven more than 60 years of computational study of materials and molecules~\cite{Jones1924,Alder1957}. 

Traditional interatomic potentials relied heavily on physically motivated functional forms and expert-driven parameterization, including Lennard-Jones interactions, bonded force fields, electrostatic partial charges, and harmonic or Fourier expansions for molecular interactions. In contrast, machine-learned interatomic potentials (MLIPs) approximate the Born-Oppenheimer potential energy surface directly from quantum mechanical reference data, typically using energies, forces, and stresses obtained from electronic structure calculations.
The topic of MLIPs is not new: the first models were reported almost two decades ago~\cite{behler2007}. However, recent advances in data availability, neural network architectures, equivariant representations, and large-scale computational infrastructure have dramatically expanded their scope and applicability. Key developments in atomic environment representations during this period include Behler-Parrinello symmetry functions, smooth overlap of atomic positions (SOAP), the atomic cluster expansion (ACE), its message-passing extension MACE, and the \(E(3)\)-equivariant framework \verb|e3nn|~\cite{behler2007,bartok2013soap,drautz2019ace,batatia_mace_2022,thomas_tensor_2018,geiger2022e3nn}. The progression of MLIP architectures is described in more detail in Sec.~\ref{better-data}.
A major recent development was the advent of ``universal'' potentials - trained on a wide and diverse dataset and capable of simulating multiple chemistries, with little or no adaptation~\cite{deng_chgnet_2023}.
These universal models can often be fine-tuned to a desired application, with very little training data, to become accurate emulators of quantum mechanical energies and forces in systems that were not well represented in the original training data. 
The community has borrowed terminology from the world of language modelling and has often dubbed these ``foundation'' models.

In this article, we explore six topics canvassed from our \texttt{ML4Atoms} reading group, which represent a subjective, but hopefully meaningful, set of open questions related to the current state-of-the-art of foundation model MLIPs. We have also sought and incorporated the input from some of the leading developers of MLIPs, to ensure that the coverage is as reflective of the current state of a fast moving field as possible. The questions are as follows:  
\begin{enumerate}
    \item \nameref{definition}
    \item \nameref{models-or-data}
    \item \nameref{long-range}
    \item \nameref{new-physics}
    \item \nameref{scaling-mlips}
    \item \nameref{benchmarks}
\end{enumerate}

We start by discussing what might define a foundation model MLIP. We borrow a taxonomy from the machine learning community that uses five categories --- expressivity, scalability, memorization, multimodality, and compositionality --- to define a foundation model. 

\begin{figure}[h!]
    \centering
    \includegraphics[width=0.9\linewidth]{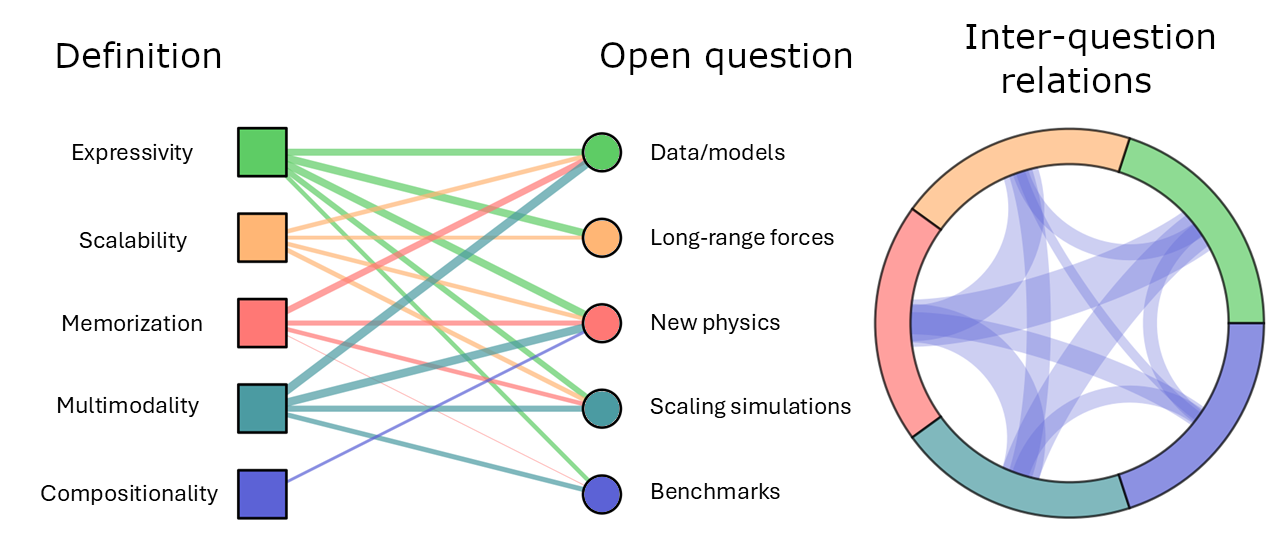}
    \caption{The connections between definitions and open questions. The text of the article has been analysed using a cosine similarity of vectorized embeddings. On the left, we show how the definition criteria link to the questions, on the right, the relations between the questions (colour-coded as on the left) are displayed. The line widths reflect the similarity of the embeddings of the content.}
    \label{fig:mlip-relations}
\end{figure}

In Figure~\ref{fig:mlip-relations} we show how these definitions relate to the five subsequent open questions, and we further show how the questions relate to one another. The subsequent discussion allows us to take a step back from the growing stream of new research in MLIPs and to consider the questions that we believe will remain at the crux of development and application moving forward in this subject.

\section{What is the minimal definition of an atomistic foundation model?\label{definition}}

%
%
Classical force fields are powerful but inherently specialized, generally requiring substantial domain expertise, hand-crafted functional forms, and parameterization to tailor them to specific chemistries or thermodynamic conditions. 
This bespoke nature makes them effective for targeted tasks but fundamentally unscalable. 
Foundation MLIPs, in contrast, aspire to transcend this fragmentation. 
The emerging class of transferable, pre-trained models, capable of supporting many downstream simulations with minimal task-specific engineering, invites comparison with the broader notion of foundation models in AI.

To ground this comparison, we draw on the framework of Bommasani et al.~\cite{bommasani_opportunities_2022}, who identified five characteristics of foundation models: expressivity, scalability, memorization, multimodality, and compositionality. 
In this section, we adapt each criterion to the context of atomistic modelling and assess how current MLIPs align with, or fall short of, these expectations. 
This domain-specific reinterpretation is not merely an exercise in taxonomy; it highlights the structural requirements any MLIP must satisfy to serve as a general-purpose basis for atomistic simulation, and helps define the axes along which atomistic foundation-model capability can meaningfully grow.
These criteria also provide the conceptual structure for the open questions that follow in the remainder of the article. 
Many of the debates surrounding MLIPs are, at their core, questions about what a foundation model ought to be. 
By examining the five criteria below, we identify the capabilities that distinguish a genuine atomistic foundation model from a highly accurate but narrow ML force field, and we set the stage for the critical issues explored in the subsequent questions. 

To emphasise the relationship of the criteria to the open questions, and the open questions to each other, we present an analysis of the contents in Fig.~\ref{fig:mlip-relations}. 
Here we have vectorized the text of the article, calculated cosine similarities between sections, and created a bipartite graph of the relations \footnote{ While we find these criteria very useful to understand MLIPs there is important distinction between atomistic foundation models and LLM-type foundation models. While MLIPs are foundational in terms of simulating chemistry, they are not direct analogues of the extremely large, highly general foundation language model. MLIPs are typically many orders of magnitude smaller than even moderate LLMs.}. We now proceed to examine how each of these criteria apply to foundational MLIPs.


    \subsection{Expressivity} 
%
 Expressivity defines the model's capacity to represent or map complex, non-linear functions from input to output.  
    Formally, if $\mathcal{F}_{\theta}$ is the family of functions parameterized by the model parameters $\theta$, then expressivity characterises how large or complex the set $\mathcal{F} = \{ f_{\theta} : \theta \in \Theta \}$ is and whether it spans the domain and codomain of the problem at hand. 
    Much of the expressivity of neural networks has been attributed to the stacking of layers, facilitating flexible function representation given sufficient data.
    For atomistic foundation models, we can cast expressivity as the set of physical interactions that an MLIP can capture.

     The expressivity of a model is fundamentally linked to the inductive biases made when designing it. Inductive biases are the assumptions that we make about the relationships in the data and that are built into the model design or training.
     One of the most common biases in MLIPs is the use of local message-passing architectures to model atomic interactions; the consequences of this bias on the ability to express long-range forces is explored in Sec.~\ref{long-range}. 

     Another factor that affects the expressivity of MLIPs is their body-order, i.e. the highest order of interatomic interactions which they can capture, where second-order interactions are bonds, third-order bond angles, and so on.  While early models were usually 2-body, 3-body and subsequently higher-order models have emerged. 
     Initially, local many-body expansions and message-passing neural networks represented distinct inductive biases: the former provided systematically improvable body-ordered descriptions of atomic environments, while the latter propagated information across atomistic graphs. Recent architectures such as MACE and GRACE \cite{batatia_mace_2022, lysogorskiy2026graph} have substantially unified these viewpoints by combining explicit equivariant many-body representations with graph-based message passing.
     While higher body order increases expressivity, in the case of rotationally equivariant neural networks, it comes at the expense of higher computational costs\footnote{ Drawing on the experience of cluster expansion simulations, body orders beyond 3 are rarely significant in crystalline solids.}.
     
     Recently, there have also been  architectures based on transformers~\cite{ kreiman2025transformersdiscovermolecularstructure}. Transformer-based models are built on the attention mechanism, which is at the core of large language models (LLMs)\cite{vaswani2017attention}. This is a different inductive bias to local message-passing routines and can more naturally model long-range forces through global attention mechanisms, overcoming the locality constraints associated with finite-range message passing. Although this flexibility can increase representational expressivity when compared to message-passing architectures, it may reduce sample efficiency when training from scratch on limited labelled data. In practice, this dichotomy captures the general trade-off between task-specialization and expressivity.  

    Understanding the practical and theoretical limits of MLIP expressivity is central to determining the domains over which foundation models can reliably generalize; it also cuts across all of the open questions discussed in this paper. In Sec.~\ref{models-or-data} we discuss the importance (or otherwise) of improving model expressivity, in Sec.~\ref{long-range} we consider a particular (and very important) case of long-range forces where there are competing reports about expressivity limitations, in Sec.~\ref{new-physics} we discuss the evidence around the ability of MLIPs to generalise beyond training data and explore new physics, and in Sec.~\ref{scaling-mlips} we discuss how expressivity and computational performance are linked.

    \subsection{Scalability} 

 In the context of foundation models, scalability is increasingly characterized through empirical scaling laws that relate model performance to key resources such as dataset size $N$, model parameters $P$, or compute $C$. These relationships are often well-described by power laws of the form:
\begin{equation}
    \mathcal{L}(N) \propto N^{-\alpha}, \quad
    \mathcal{L}(P) \propto P^{-\beta}, \quad
    \mathcal{L}(C) \propto C^{-\gamma},
\end{equation}
where $\mathcal{L}$ denotes a suitable loss or error metric, and the exponents $\alpha$, $\beta$, and $\gamma$ quantify the efficiency with which performance improves as each resource is increased. Larger exponents correspond to more favourable scaling behaviour.

  In this sense scalability is also related to expressivity, in that the expressivity of a given architecture sets upper bounds on the scalability of a model: for example, if a model has only one layer of message passing and therefore each atom sees only it's nearest neighbours, no amount of training data will facilitate accurate capturing of higher-order interactions to overcome this representational limitation. In contrast, if higher-order interactions can be captured implicitly, increased capacity/data could improve performance.

    Another consideration for scaling is that models should be easy to train on, and predictably improve, with increasingly large datasets. Foundation models should also be possible to fine-tune without losing their original performance; see the discussion on memorization for more on this. Foundation models should also be practically efficient, compatible with common computer architectures and able to take advantage of parallelization. These aspects are considered in more depth in Sec.~\ref{scaling-mlips}.
    To some extent, these two goals -- training on big datasets and efficiency in inference -- are in opposition: Increasing model capacity and representational complexity typically improves expressivity at the expense of computational efficiency. The notion of scalability must, therefore, consider both. New research trying to balance these competing requirements is discussed in detail in Section~\ref{scaling-mlips}.

    \subsection{Memorization} 
    Traditional foundation models require ``[knowledge encompassing] both a broad understanding of the world as well as specific mastery of niche subjects or particular facts.''\cite{bommasani_opportunities_2022}. 
    
    For atomistic foundation models, the analogue of memorization is less straightforward than in language models, where factual recall can often be directly probed. In the MLIP setting, a more useful interpretation concerns retention and adaptation: to what extent can a model acquire new domain-specific knowledge through fine-tuning while preserving previously learned capabilities? These questions are related to multimodality, discussed below, and are also considered in depth in Sec.~\ref{models-or-data}, where the role and nature of the training data are considered. The question of memorization versus generalization of learning is also fundamentally linked to the ability of MLIPs to explore new physics and is discussed in Sec.~\ref{new-physics}.
    \subsection{Multimodality} 
    In the context of foundation models, multimodality refers to the ability of a single model to process, represent, and generate information across multiple data modalities, such as text, audio, video, and images. 
    Multimodality may seem less immediately applicable to MLIP foundation models. A truly multimodal machine learning model for chemistry might predict spectroscopic data, or even microscopy images, given an input structure. Property prediction models are also often trained on multiple properties at the same time.~\cite{devi2024optimal} However, this is beyond the realm of the prevailing understanding of foundation MLIPs. We therefore reinterpret this criterion in the context of atomistic modelling as multi-fidelity learning.

    Typically, MLIP models have been trained on single sources of data, with a consistent generating process (usually density functional theory within a particular ansatz with consistent calculation settings).
    This kind of setting results in a model which is self-consistent for the given training setup, but may limit transferability across chemical domains or simulation fidelities. Although differences in electronic-structure fidelity do not constitute distinct modalities in the strict machine-learning sense, they do represent heterogeneous sources of supervision with differing resolutions, approximations, and error characteristics. 

    Multi-fidelity and multi-property learning of atomistic properties in non-MLIP models (targeted at properties other than energy and force) has demonstrated improved cross-fidelity learning and models that generalize better to new data~\cite{chen2021learning, devi2024optimal}. 
    Recent developments~\cite{Kim2024MultifidelityMLIP,batatia2025mh} in fine-tuning for MLIPs have also introduced ``multi-head'' fine-tuning allowing for the introduction of data from new fidelities/generating processes, without catastrophically degrading performance on previously learned domains.
    Recent work has also looked at fine-tuning of MLIPs~\cite{gumber_going_2025} using multi-modal (in the traditional sense) data. These developments and questions around training data are discussed more in Sec.~\ref{models-or-data} and Sec.~\ref{new-physics}.

    \subsection{Compositionality}
   Bommasani et al. defined compositionality as relating to the modularity of the training data, the learned representation, or the models themselves. However, we reinterpret compositionality for atomistic simulations as a model's ability to accurately describe large systems that are composed of smaller units with which the model is more familiar. For instance, a truly foundational atomistic model should be able to accurately describe polymers having only encountered the local atomic environments associated with the constituent monomers during training. This property can in part stem from the design choice of predicting local atomic energies which sum to yield the total energy. However, truly satisfying compositionality for generic systems will likely require augmentation of the local energy with explicit treatment of long-range interactions (see Sec.~\ref{long-range}).

An alternative perspective on compositionality, perhaps closer to the original foundation model sense, is whether the internal representations learned by MLIPs can be composed with or transferred to downstream tasks beyond energy and force prediction. Recent work suggests this may already be occurring: last-layer features from PET-MAD serve as effective collective variables for phase transition detection and as general-purpose structural descriptors for `materials cartography'~\cite{mazitov_massive_2025}, while collective variables for protein folding have been extracted from frozen biomolecular foundation model representations~\cite{psyga2026}. Systematic comparison of latent features across universal MLIPs reveals that different architectures encode chemical space in distinct but partially reconstructible ways~\cite{ctmhcc2026}, and cross-modality alignment studies show that scientific foundation models — including MLIPs, string-based models, and even LLMs — are converging toward shared representations of matter as they improve~\cite{eylg2025}.

The criteria outlined in this section suggest that current MLIPs satisfy different subsets of the capabilities we associate with atomistic foundation models, rather than realizing all of them simultaneously. However, we have intentionally framed this section as an open question: there will be differing opinions on what defines a foundation model, and it remains unclear whether we \textit{need} every MLIP to be truly foundational. Nevertheless, we expect these dimensions to define many of the key directions along which state-of-the-art MLIPs will continue to evolve.

\section{Do we need more data, better data, or better models?\label{models-or-data}}

A central tension in the development of MLIPs echoes a long-standing debate in AI: should progress come primarily from carefully designed inductive biases or from scaling models and data? Sutton’s ``Bitter Lesson''~\cite{sutton2019bitter} argues that, in the long run, systems that rely less on human-engineered structure and more on compute-enabled general methods tend to outperform those with strong built-in priors. In contrast, recent responses (the ``Bittersweet Lesson'' and related arguments~\cite{kranmar2025bittersweet}) contend that inductive biases remain essential in domains grounded in physics, where symmetries, conservation laws, sparsity, and locality are not heuristics but truths about the world.

This raises a key question for MLIPs: where should we place our research investment? Should we engineer better architectures that embed more physics, or simply increase model capacity and trust optimization to learn the right behaviours from data? Should we collect dramatically larger and more diverse datasets, or instead curate smaller, higher-fidelity corpora targeted to the chemical physics of interest? And, crucially, how do model complexity, data volume, data quality, and inductive bias interact? We consider these questions and related evidence from the perspectives of better models, more data, and better data.



\subsection{Better models}

MLIPs have progressed by incorporating the right \emph{inductive biases}. 
Early models enforced permutation and translation invariance via atom-wise energy decompositions and symmetry functions, enabling a single network to handle varying sizes and compositions~\cite{behler2007}. 
Symmetry-respecting descriptors (e.g.\ SOAP) were unified by ACE into a complete, linearly scaling basis for local environments~\cite{bartok2013soap,drautz2019ace}, making potential energy surface (PES) learning faithful and reusable across chemistry.

A key step was encoding \(E(3)\) \emph{equivariance} inside the network.
Equivariant GNNs (EGNNs) such as NequIP and MACE~\cite{batzner20223,batatia_mace_2022} propagate vector and tensor features that rotate with geometry, improving data efficiency (Fig.~\ref{fig:batzner_plot}) and transferability --- especially with force/stress supervision. See Sec.~\ref{scaling-mlips} for a full explanation of equivariance in MLIPs. However, when one controls for the increase in parameter count incurred by adding equivariance, the improved efficiencies are often negated\cite{NEURIPS2024_fad8e191}.
Equivariant features can reduce sample complexity and stabilise directional chemistry, but when physics demands symmetry breaking or anisotropy, recent methods relax or adapt equivariance accordingly~\cite{smidt2021,xie2024,hofgard2024}. 
Mathematically, conservative models need not be internally equivariant: an invariant energy yields properly transforming derivatives. 
Yet non-equivariant, non-conservative models (e.g.\ the diffusion-based Orb-v3 direct model~\cite{rhodes_orb-v3_2025}) can match EGNNs on prediction of higher-order PES-derived properties without the computational overhead of many extra automatic differentiations~\cite{riebesell2025framework}. 
However, direct-force heads risk non-conservative fields that undermine MD stability and phonon accuracy; where physical consistency is paramount, energy-based (conservative) models are preferred~\cite{bigi2025darkside,loew2025}. 
Despite this, direct-property predictors may be augmented by conservative models to accelerate training whilst retaining robust models, provided sufficient high-quality data is used~\cite{ rhodes_orb-v3_2025,bigi2025darkside}.

Building equivariance into a model versus learning equivariance from data is still an ongoing debate~\cite{nr2025,bbhc2025}. 
The arguments that favor building in equivariance are, first and foremost, physical consistency, and also improved data efficiency, particularly in the small-dataset regime~\cite{batzner20223}.
On the other hand, models learning equivariance from the data may offer advantages in computational cost and scalability to very large datasets and model sizes~\cite{qu2024importance,rhodes_orb-v3_2025,mazitov_pet-mad_2025,qu_recipe_2026}.
In addition, it remains unclear whether enforcing equivariance constraints introduces additional challenges for optimization, although clear evidence for such effects in realistic models is still limited~\cite{xie2025tale}.

Speed and cost gains now also come from the training stack: \texttt{e3nn} standardized $SO(3)$ tensor operations and \texttt{cuEquivariance} and \texttt{OpenEquivariance} fused them into high-performance CUDA kernels; compiler/ahead-of-time tooling (e.g.\ \texttt{torch.compile}, \texttt{torch.jit}) and distributed parallelization further raised throughput without changing the physics~\cite{geiger2022e3nn,cueq2024,openequivariance,paszke2019pytorch,pytorch_torchscript_docs,han_distmlip_2025}. 
Optimiser choices have also matured: using AdamW over Adam, or newer options such as Muon, reduce training steps; and cosine/OneCycle schedules with EMA further provide robust baselines~\cite{loshchilov2019adamw,jordan2024muon_blog}.

Foundation uMLIPs (M3GNet, CHGNet) pre-trained on large Materials Project (MP) corpora established a template for robust relaxations and MD across chemistry~\cite{chen2022m3gnet,deng_chgnet_2023}. 
\emph{MACE–MP–0} extended this with an \(E(3)\) backbone and public checkpoints (e.g.\ MP–0b), and fine-tuned variants now frequently achieve state-of-the-art benchmark performance (see Sec.~\ref{benchmarks} for detailed discussion of benchmarks). 
Transfer learning (fine-tuning, multi-/mixed-fidelity) reliably beats from-scratch trained models at equal compute; multi-head training broadens property coverage and provides useful uncertainty estimates~\cite{merchant_scaling_2023,radova_fine-tuning_2025,messerly2025,batatia2025mh}. 
Notably, compact equivariant models (e.g.\ \emph{Nequix}, $\sim$700k parameters) have emerged that rival much larger stacks, when paired with modern kernels and recipes, on the Matbench discovery benchmark~\cite{koker2025nequix}.

\subsection{More data}

\begin{figure}
    \centering
    \includegraphics[width=0.8\linewidth]{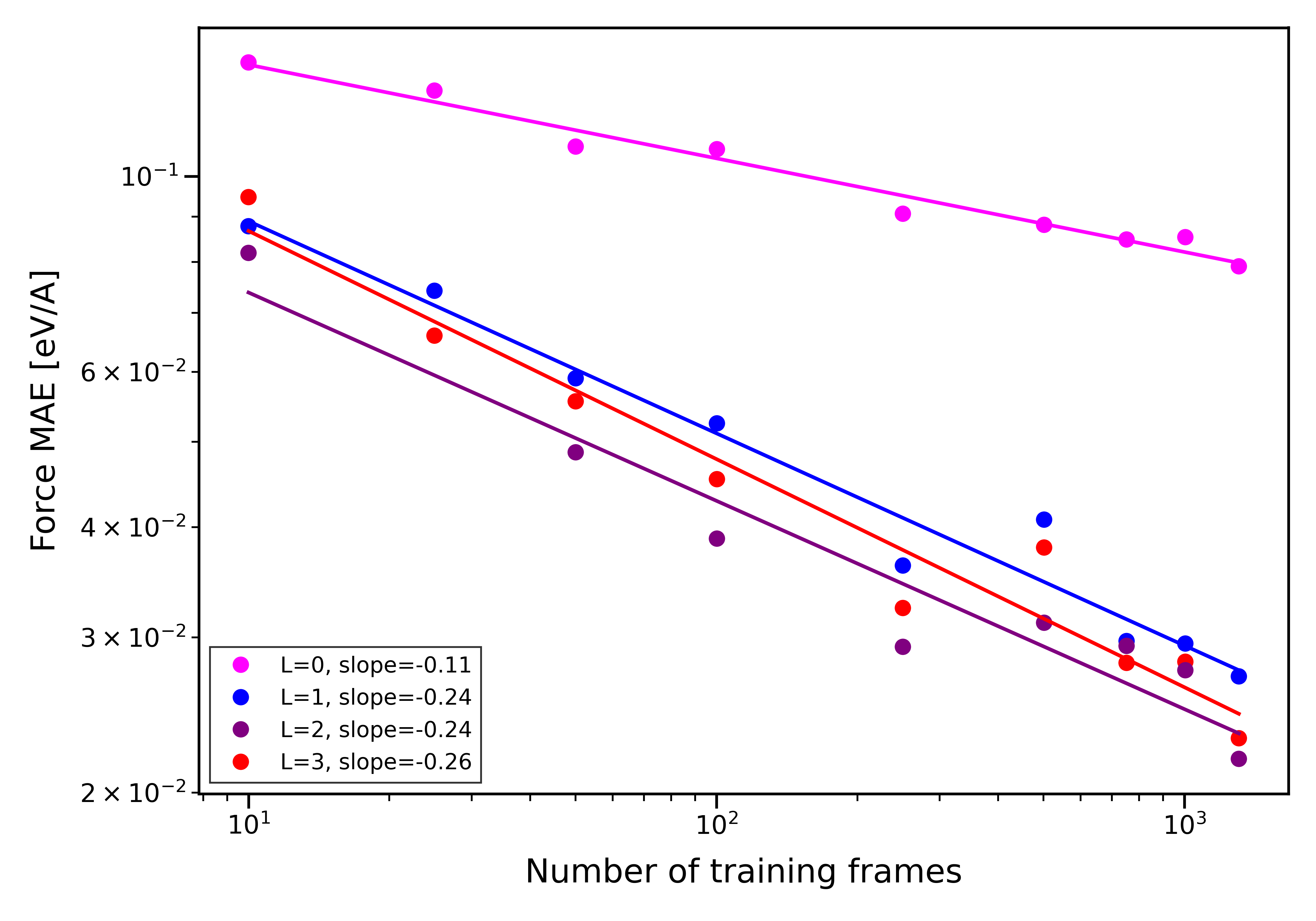}
    \caption{Log-log plot of the predictive error on the water data set from~\cite{bingqingAbInitioThermodynamics} using NequIP with rotation order $L \in \{0,1,2,3\} $ as a function of training set size, measured via the force MAE. Figure from Batzner~\textit{et al.}~\cite{batzner20223}}
    \label{fig:batzner_plot}
\end{figure}

As a general trend, state-of-the-art MLIPs show strong positive correlations between performance and dataset size,~\cite{vonlilienfeldRetrospectiveDecadeMachine2020} as highlighted by the MatBench discovery~\cite{riebesell2025framework}
scaling results for GNoME, Mattersim, Alexandria, and OMat24~\cite{merchant_scaling_2023, yangMatterSimDeepLearning2024, schmidtImprovingMachinelearningModels2024, BarrosoLuque2024OMat24}.

In recent work with the MatPES data-efficient sampling scheme~\cite{kaplanFoundationalPotentialEnergy2025}, the authors show that MLIPs trained on a compact, representative ~400K dataset (curated via 2DIRECT~\cite{qi_robust_2024}) matched or exceeded models trained on the 250$\times$ larger OMat24~\cite{BarrosoLuque2024OMat24} dataset. The Massive Atomic Diversity (MAD) dataset similarly captures data for inorganic materials and molecules, with an emphasis on diversity over scale.\cite{mazitov2025massive}
These examples suggest brute-force scaling of data volume is often computationally inefficient, likely because current high-throughput datasets have significant systematic and unsystematic noise from mixed PBE/PBE+U calculations, different computational settings used in DFT, and they contain mainly near-equilibrium structures, which do not accurately inform on global PES shape. 
Building on this point, the benefits of increased data diversity are tied to the complexity or expressivity of the model. Schmidt et al.\cite{schmidtImprovingMachinelearningModels2024}. show how models with insufficient complexity have performance saturation at high dataset size regimes. 
The paper shows, for example, how the performance of ALIGNN~\cite{choudharyAtomisticLineGraph2021} plateaus on the Alexandria dataset compared to more complex MLIPs like MACE~\cite{batatia_mace_2022} and M3GNet~\cite{chen2022m3gnet}, which do not. 

Additionally, increasing data volume without scaling model complexity can degrade generalization\cite{gibsonWhenMoreData2024}. 
This may arise when a model lacks the capacity to capture the increased variance in larger, more diverse datasets. 
Figure~\ref{fig:batzner_plot} shows how scaling model complexity by increasing expressivity (equivariance and higher order tensor features introduced from $l \leq 1$) offers not only performance gains for NequIP~\cite{batzner20223}, but also improved data efficiency and convergence rates relative to the invariant ($l = 0$) baselines.

The case for over-parameterization is supported by the `deep double descent' phenomenon observed in fundamental deep learning research\cite{nakkiranDeepDoubleDescent2019}. Empirical evidence suggests that test set error peaks at an `interpolation threshold', where model parameters are approximately equal to data size. However, increasing model capacity beyond this threshold into the over-parameterized regime initiates a secondary decline in test error. The authors propose that in this regime the models are capable of `absorbing' the noise while maintaining generalization. This may be particularly relevant for atomistic modelling, where the intrinsic approximation errors of the PBE functional introduce substantial systematic noise into heterogeneous DFT datasets.

Additionally, research on scaling laws for neural language models suggest that model parameters, dataset size, and compute should all be scaled in tandem to maximise performance gains\cite{kaplanScalingLawsNeural2020}. 
While studies on general deep learning models may be useful to guide scaling, the physics priors present in MLIPs (eg. equivariance) may change the learning landscape by reducing the burden on the model to implicitly learn physical laws from data\cite{frey2023scaling}. However, robust empirical scaling laws analogous to those observed in language models remain comparatively under-explored for MLIPs.

Together, these show that the relationship between data and performance goes beyond sheer dataset size. Performance gains require balancing data volume, diversity, model complexity and the strength of incorporated priors to the model. 
Prioritising model complexity alongside efficient and informed dataset curation appears more important than relying on brute-force scaling of datasets.

\subsection{Better data \label{better-data}}

High-throughput density functional theory (HT-DFT) has accelerated materials discovery, but it is not without limitations. 
Recent work~\cite{Kuryla2025DFTForces} highlights that the energy and force criteria used in HT-DFT datasets, such as those from the MP database, are often too lenient, leading to inaccuracies in derived properties. 
These limitations become particularly pronounced when training MLIPs, which rely heavily on the fidelity of input data.
Dataset generation is typically the most computationally expensive component of the MLIP development pipeline. 
This practical constraint places a premium on data efficiency, motivating approaches that prioritise representative and information-rich configurations over brute-force increases in dataset size.

While generalized gradient approximation (GGA) functionals like PBE~\cite{perdew_PBE_1996} are sufficient for many bulk property predictions, they fall short in capturing defect energetics \cite{Meggiolaro2018} and phonon properties\cite{Taheri2018}. 
Hybrid functionals (B3LYP~\cite{stephens_b3lyp_1994}, HSE06~\cite{krukau_hse_2006}) offer improved accuracy but at significantly higher computational cost. Deng \textit{et al.}~\cite{deng_systematic_2025} demonstrated that MLIPs trained solely on PBE/GGA for equilibrium geometries data suffer from systematic softening, particularly in high-energy or distorted atomic environments. 
This results in underestimation of defect formation energies and poor transferability to out-of-distribution configurations. 
A critical challenge for universal MLIPs is the accurate prediction of vibrational properties, which they often underestimate due to overly soft potential energy surfaces~\cite{deng_systematic_2025}. This issue originates from training primarily on near-equilibrium structures, a common strategy in datasets like sAlex and MPTrj~\cite{deng_chgnet_2023, Fu2025SmoothExpressive}. 
The solution lies in data quality and diversity. 
For instance, the OMat24 dataset~\cite{BarrosoLuque2024OMat24} resolved this softening issue by incorporating diverse, high-energy, non-equilibrium configurations generated through rattled sampling or \textit{ab initio} molecular dynamics runs.
It is also important to distinguish between structural diversity and diversity of local atomic environments. MLIPs operate on local environments, and their ability to generalise is primarily governed by the diversity of these environments rather than the diversity of global structures. 
This distinction is often overlooked, leading to concerns about transferability between, for example, crystalline and amorphous systems. However, many local coordination motifs are shared across such systems, suggesting that the effective space of relevant environments may be more constrained than is sometimes assumed.

The computational expense of generating high-fidelity data is a major constraint in developing machine-learned interatomic potentials. 
A promising solution is the use of multi-fidelity datasets.
Transfer learning and multi-fidelity approaches offer promising solutions to the high cost of generating hybrid-level data. 
Recent studies~\cite{Ko2025HighFidelityML} demonstrated that models trained on a 90:10 mix of PBE:SCAN functional data can match the accuracy of models using eight times more pure SCAN data. 
This principle can be extended through more advanced multi-fidelity schemes.

Furthermore, not all properties require high fidelity; using high-fidelity energies with low-fidelity forces can yield performance nearly as good as using both at high fidelity. 
This principle extends to advanced methods, at least at small data scales, where models learn to map between fidelities, achieving a 10$\times$ reduction in the need for costly hybrid DFT calculations for defect properties~\cite{messerly2025, Kim2024MultifidelityMLIP}. However, it remains to be seen if these conclusions hold at larger data scales.
Further research has shown that by building models that actively learn the mappings between data fidelities, it is possible to integrate diverse data ranging from standard DFT and hybrid functionals to experimental data. 
This intelligent integration is particularly useful for costly properties like defect formation energies, where it can reduce the need for high-fidelity calculations by an order of magnitude~\cite{Oerder2025MultiFidelity}.

A complementary frontier is the propagation of uncertainties inherent in the DFT calculations themselves. Recent work on differentiable DFT frameworks~\cite{sph2025} enables the estimation and propagation of uncertainties due to convergence settings to computed energies, as well as downstream observables such as lattice constants or band structures. Such per-sample DFT error bars could inform multi-fidelity training by allowing heterogeneous data to be weighted by its estimated reliability, rather than relying on uniform loss weighting across datasets of varying fidelity.

Taken together, current evidence suggests that progress in MLIPs is unlikely to arise from scaling any single axis in isolation. Increasing dataset size alone does not guarantee improved transferability or physical fidelity, particularly when model expressivity or data quality become limiting factors. Likewise, increasingly sophisticated architectures cannot compensate indefinitely for incomplete or noisy supervision. Instead, the emerging picture resembles modern foundation-model scaling more broadly: performance improvements arise from coordinated advances in model design, dataset diversity and fidelity, optimization, and computational infrastructure. Determining the relative importance of these factors, and whether physically informed inductive biases remain advantageous at extreme scale, remains one of the central open questions in atomistic foundation models.

\section{Can MLIPs really handle long-range interactions, and does it matter?\label{long-range} }

As discussed previously, one approach to generating better models is to include long-range interactions. 
%
The locality of interactions in MLIPs is in part justified by Kohn's `near-sightedness' principle~\cite{Kohn1996,Prodan2005}. However, there are many different systems~\cite{ grisafi2019incorporating,yue2021short, ko2023accurate, ko2021general, staacke2021role, cheng2025latent} where it has been shown in the literature that it is beneficial to include long-range interactions to fully describe the physics of the system:
notably, Grisafi and Ceriotti~\cite{grisafi2019incorporating} showed that the inclusion of long-range interactions is required to accurately capture the dissociation curves of charged dimers by machine learning models. 
Meanwhile, Yue~\textit{et al.}~\cite{yue2021short} showed that, despite the fact that local machine learning models are sufficient to obtain accurate predictions in the liquid phase, the short-range nature of machine learning models fails to describe the properties of the clusters and vapour phase. 
Similarly, Niblett, Galib, and Limmer~\cite{Niblett2021} showed that short-range machine-learning models can yield a bias in the orientational order profile, which can be alleviated by supplementing the model with a Coulomb baseline.
One common theme here is inhomogeneity~\cite{janecek2006long}:  systems with interfaces or anisotropy are more sensitive to long-range forces than their homogeneous congeners. 
Additionally, Behler~\textit{et al.}~\cite{ko2021general} have shown that systems where the partial charges/charges change during the simulation require both the inclusion of long-range interactions and charge equilibration, for example, in describing NaCl clusters~\cite{ko2023accurate} and gold binding to MgO~\cite{ko2021general}.
However, recent work~\cite{Rumiantsev2025} has shown that these systems can also be solved with sufficiently expressive MLIPs without explicit charge equilibration: Long-range models like SpookyNet~\cite{unke2021spookynet}, CACE-LES~\cite{cheng2025latent}, or LOREM~\cite{Rumiantsev2025} pass this benchmark, and, given an appropriate effective cutoff, so do short-range models like MACE~\cite{batatia_mace_2022} and PET~\cite{pc2023}. Additionally, data-driven approaches to incorporate long-range interactions have also been explored through combining a local neighbourhood attention with an all-to-all attention mechanism with no radius cutoff, with the ability to scale to training on $\mathcal{O}(100\text{M})$ data samples~\cite{qu_recipe_2026}. This is a domain with similar questions of whether this information should be encoded explicitly, or if some of this information can be learned from the data.
The cumulene chain example~\cite{ucsgpstm2021} illustrates a qualitatively different class of long-range problem arising from electronic delocalization, rather than classical electrostatics. This benchmark, since adopted by several subsequent studies~\cite{frank2023so3krates,Rumiantsev2025,batatia2024equivariantmatrixfunctionneural}, highlights that long-range in MLIPs is not synonymous with electrostatics.
Describing charge defects~\cite{staacke2021role, freysoldt2011electrostatic, french2010long} in solids also requires the inclusion of long-range interactions. 
Additionally, Berry-phase polarization and dielectric response, and longitudinal-optical versus transverse-optical (LO-TO) phonon splitting all require coherent long-range electrostatics~\cite{Resta1994, GonzeLee1997}. 
In excited states, electron–hole attraction forms excitons whose binding hinges on long-range screening~\cite{ho2008effect}.

\subsection{What are long-range interactions?}

From the above discussion, it is clear that the inclusion of long-range interactions is needed to describe the physics of many systems. 
There are many different ways to define long-range interactions, which broadly fall into two categories: those based on physics and those based on the graph/architecture of the system. 

\subsubsection{Physics based definitions}

``Long-range’’ means a local perturbation remains influential far away. 
Classically, this is set by the tail of the pair potential
\begin{equation}
  V(r)\propto r^{-\alpha},
\end{equation}
with separation $r$ and decay exponent $\alpha$. 
In spatial dimension $d$, $\alpha<d$ gives \emph{strong} long-range (distant particles contribute non-negligibly); $d<\alpha<\alpha^*$ (with $\alpha^*$ some critical exponent) gives \emph{weak} long-range~\cite{DefenuRMP2023}. 
Canonical examples are Coulomb ($1/r$), dipole–dipole ($1/r^3$), and dispersion ($C_6/r^6$), although a further distinction can be made. 
In periodic systems, Coulomb interactions between point charges cannot be converged by simply imposing a real-space distance cutoff, while for dipole–dipole and dispersion interactions, a sufficiently large real-space cutoff can, in principle, yield a converged result. 
%
%
Dimensionality and environment can also reshape the tail of the interactions, e.g.\ electrostatic screening, and the Rytova–Keldysh form in two-dimensional layers~\cite{Keldysh1979}.
In condensed phases, for example, the $r^{-\alpha}$ tails control macroscopic polarization as the energy gradient with an electric field, and produce longitudinal-optical versus transverse-optical (LO-TO) phonon splitting through dipole–dipole couplings (i.e. Born effective charges and the high-frequency dielectric tensor)~\cite{Resta1994, GonzeLee1997}.

\medskip
\noindent In quantum chemistry, long-range also refers to electron correlation that couples \emph{distant} charge fluctuations. 
For two well-separated fragments, the dispersion interaction decays as $r^{-6}$, and higher-order and many-body terms arise from fluctuating dipoles. Non-local correlation methods (e.g.\ vdW-DF functionals) recover these asymptotics in practice.



An alternative definition from a statistical physics perspective, uses the correlation definition from the connected two-point function for a scalar field $\phi$, $C(r)=\langle \phi(0)\phi(r)\rangle-\langle \phi\rangle^2$. 
Away from criticality $C(r)\sim e^{-r/\xi}$, so the \emph{correlation length} $\xi$ sets the range of the correlations. At continuous phase transitions $\xi\!\to\!\infty$ and $C(r)\sim r^{-(d-2+\eta)}$ (scale-free, effectively long-range) with $\eta$ the critical exponent~\cite{Cardy1996}.






\subsubsection{Graph and network definitions}
In atomistic graphs, nodes (atoms) and edges (interactions) define a \emph{topological} distance $d_G(u,v)$ (e.g.\ shortest-path, resistance/commute-time, or diffusion). Here, long-range means that the prediction at node $u$ depends materially on nodes $v$ with large $d_G(u,v)$
, even when Euclidean separations are small (or vice versa).
To move beyond a purely graph-distance-based notion of long-range interactions, one can instead consider the \emph{influence} of one node on another within the neural-network representation~\cite{Bamberger2025}. Defining the influence of node \(v\) on node \(u\) as
\[
I_u(v)=\left|\frac{\partial (F(X))_u}{\partial x_v}\right|,
\]
where \(X\) is the feature vector and \(F(X)\) is the mapping induced by the neural-network architecture, one may then construct an influence-weighted average graph distance

\[
\rho_u(F)=\frac{\sum_v I_u(v)\,d_G(u,v)}{\sum_v I_u(v)}.
\]
This provides a notion of effective interaction range that depends not only on graph connectivity, but also on how strongly information propagates through the learned representation.
%


\medskip
\noindent Across all these perspectives, long-range is how far a perturbation or fluctuation meaningfully propagates; encompassing potential tails, correlation length, global responses, and graph-theoretic influence.

\subsection{Why do models fail to capture long-range interactions?}
Message Passing Neural Networks (MPNNs), one of the most widely used architectures for MLIPs, exhibit a local inductive bias~\cite{giovanni2023on}, as message aggregation for each node is typically limited to its neighbourhood.
This locality reflects the assumption that short-range interactions (up to a large enough receptive field) govern the physics, while taking into account the limitations due to computing costs: MPNNs with a fixed cutoff scale linearly with system size~\cite{balcilar2021breakinglimitsmessagepassing}, whereas, for example, transformer-based models with global attention scale quadratically in the standard softmax formulation~\cite{cai_connection_2023}, though there is a large body of work on subquadratic alternatives, for example linear attention~\cite{kvpf2020}, structured state-space models~\cite{ggr2021,gd2024}, and Euclidean fast attention~\cite{fcmu2024} for atomistic models.
However, this locality bias of MPNNs limits expressiveness~\cite{xu2019powerfulgraphneuralnetworks, kreiman2026understanding} and explicitly neglects higher-order and long-range interactions. 
Consequently, MPNNs suffer from three related issues: over-smoothing, where node embeddings become indistinguishable with depth~\cite{li2018deeperinsightsgraphconvolutional,oono2021graphneuralnetworksexponentially}; over-squashing, where exponentially growing information is compressed into fixed-size representations~\cite{alon2021bottleneckgraphneuralnetworks,topping2022understandingoversquashingbottlenecksgraphs}; and under-reaching, the need for at least $K$ layers to propagate information across $K$-hop neighborhoods~\cite{alon2021bottleneckgraphneuralnetworks}.
While over-smoothing, over-squashing, and the related problem of under-reaching are widely recognised as sources of poor long-range performance in MPNNs~\cite{giovanni2023on}, their definitions and explanatory power are more nuanced. Arnaiz-Rodriguez and Errica~\cite{arnaizrodriguez2025oversmoothing}, for example, show that over-smoothing is not inherent to all deep graph networks and argue that over-squashing can arise through two distinct mechanisms - a computational tree~\cite{alon2021bottleneckgraphneuralnetworks}, and/or a topological bottleneck~\cite{digiovanni2024doesoversquashingaffectpower} - both of which have recently been investigated by H. Blayney \textit{et al.}~\cite{blayney2025glstm}. Furthermore, A. Arroyo \textit{et al.}~\cite{arroyo2025vanishinggradient} highlight the connection between over-smoothing, over-squashing, and the vanishing gradient problem known from recurrent neural networks.
 
\subsection{Approach to mitigate problems with long-range interactions} 

Numerous approaches have been proposed to mitigate the problems with long-range interactions. Like with the interactions themselves, methods to mitigate the problems with long-range interactions can be classified into graph/architecture and explicit physics-based approaches.

\subsubsection{Graph/Architecture approaches}

One approach to mitigate the problems mentioned in the previous section is to effectively shorten the distance for message passing. What all these different methods of effective shortening have in common is that their underlying mechanisms are not always fully understood, particularly regarding their effectiveness in large-scale model architectures and with extensive datasets. One example of such an approach is the use of spatial~\cite{gutteridge2023drew} or spectral~\cite{karhadkar2023fosrfirstorderspectralrewiring} rewiring techniques to connect nodes by the newly introduced edges. The other common solution is to introduce additional nodes, such as virtual/master/supernodes~\cite{gilmer_neural_nodate, scarselli2008graph} that are either fully connected to all other nodes~\cite{cai_connection_2023}~\cite{southern_understanding_2025} or to subgraphs~\cite{hwang_analysis_nodate} and function as mediators to pass the messages between distant nodes. It has empirically been shown that introducing virtual node(s) can enhance the predictive performance of models~\cite{sestak_vn-egnn_2024,southern_understanding_2025,liu_boosting_2022} and that in fact they can approximate self-attention layers of graph transformers~\cite{cai_connection_2023} with lower memory cost~\cite{southern_understanding_2025}. Li \textit{et al.} further extended this idea by introducing neural nodes, which are essentially multiple virtual nodes that interact with one another~\cite{li_neural_2024}. 

Arroyo \textit{et al.}~\cite{arroyo2025vanishinggradient} link oversmoothing and oversquashing to vanishing and exploding gradients, proposing a state-space reformulation of GNNs to mitigate them. Earlier, Kiani \textit{et al.}~\cite{kiani2024unitary} addressed the same issue by drawing on RNN literature and introducing unitary graph convolutions~\cite{kiani2024unitary}.

Another approach is hierarchical (graph) learning methods, which mainly differ in how they construct multi-level graph abstractions and how these levels interact with each other. Construction strategies range from chemically motivated coarse-graining, such as BRICS fragmentation~\cite{li2023long}, to graph-theoretic approaches like METIS partitioning~\cite{mathys2025learn} and learned junction-tree structures~\cite{fey2020hierarchical}. Once atom-level, motif-level, and/or global graphs are defined, models vary in their cross-level message passing.  Li \textit{et al.}~\cite{li2023long} build separate architectures at the atomic and motif level, and then combine their respective final predictions to an overall one. By contrast, Sun \textit{et al.}~\cite{sun2025molgraph} combine GNNs with xLSTMs and a mixture-of-experts approach to coordinate cross-level communication even before the final prediction. Likewise, Han \textit{et al.}’s HimGNN~\cite{Han2023HimGNNAN} leverages atom–motif correspondences through transformer-based local augmentation, while Mathys \textit{et al.}~\cite{mathys2025learn} employ adaptive random walks to facilitate information flow across hierarchical layers.

In addition to this, spectral approaches have been developed, such as SOG net~\cite{ji2025machine}, which uses an efficient Fourier convolution layer to incorporate long-range effects, or ARMA using rational filters~\cite{Bianchi_2021}. Based on these I. Batatia \textit{et al.}~\cite{batatia2024equivariantmatrixfunctionneural} proposed Matrix Function Neural Networks that modify spectral approaches by modelling non-local interactions through analytic matrix equivariant functions. Several methods that use the attention mechanism have also been developed, such as Molformer, SpookyNet, and the Equiformer family, EquiformerV1 and EquiformerV2~\cite{wu2023molformer,unke2021spookynet,equiformerv1,equiformerv2}. 
The original So3krates model~\cite{frank2023so3krates} used message-passing in an auxiliary space of spherical harmonics features to model long-range interactions. However, this depends on initial features being close enough for gradient flow. Thus, this feature was removed from later versions of So3krates~\cite{fumc2024} and later long-range work focused on geometric fast attention (EFA~\cite{fcmu2024}) and explicit physical terms (SO3LR~\cite{kabylda_molecular_2025}).
The advantages of adding explicit physical terms to the model will be discussed next.

\subsubsection{Explicit Physics}
An alternative to the approaches based on the machine learning architecture is to use insight into the physics of the problem, learnt in part by classical force field development. Behler and others have classified these as third- and fourth-generation machine learning potentials~\cite{ko2023accurate, behler2021machine, kulichenko_data_2024}, with third-generation machine learning potentials capturing long-range electrostatic/van der Waals forces and fourth-generation allowing for charge equilibration~\cite{ko2023accurate, behler2021machine}.


%
Examples of early architectures inspired by classical forces include supplementing a short-range model with a fixed-partial-charge coulomb baseline~\cite{Bartk2010}, followed by models that predict position-dependent partial charges which supplement a local model with long-range interactions~\cite{Artrith2011}. 
Subsequently, the LODE representation was proposed by A. Grisafi and M. Ceriotti \cite{grisafi2019incorporating}, where an Ewald summation (see e.g.~\cite{Toukmaji1996}) is used to compute a Coulomb-like atom-density potential. This potential is projected onto an atom-centered spherical basis to yield long-range features serving as inputs to a regression model.

Recently, the latent Ewald summation (LES) method was developed~\cite{cheng2025latent}, which represents the total energy as a short-range local model plus a long-range term computed by an Ewald summation, allowing to predict electrical response properties such as polarization and Born effective charge tensors just from learning energies and forces ~\cite{King2025Machine, zhong2025machine}.

An advantage of the LES approach is its interoperability with existing short-range models such as MACE, NequIP, CACE, UMA, and CHGNet~\cite{kim2025universal}. 
In addition to that, Loche~\textit{et al.}~\cite{loche2025fast} released differentiable PyTorch and (experimental) JAX implementations of Ewald summation and particle-mesh variants (PME/P3M), further improving integration into existing frameworks. 
%
%
%

%
Building on this idea, Rumiantsev~\textit{et al.} introduced LOREM~\cite{Rumiantsev2025}, which augments an equivariant short-range message-passing model to predict short-range energies and charge-like tensors, with a global long-range message-passing block that computes equivariant atomic potential-like tensors. 
%
%
Finally, Ramasubramanian~\textit{et al.}~\cite{Ramasubramanian2025} go beyond a fixed Coulomb-like functional form for the long-range energy by expressing the long-range Ewald term as attention over reciprocal-space features.

In recent years, a number of methods have been adopted in so-called fourth-generation machine learning models which allow for charge equilibration. In each approach, the energy of the system is broken down into a local energy and a longer-range electrostatic contribution, with charges in each step determined self-consistently.  Celli~\cite{fuchs_learning_2025}, CENT~\cite{khajehpasha2022cent2}, BPopNN~\cite{xie2020incorporating} and QET~\cite{ko2025fast} use electronegativities and hardnesses, and species-dependent charge radii to predict partial charges and the Coulombic potential to allow charges to be equilibrated in the simulation. By contrast, Thomas~\textit{et al.}~\cite{thomas2025self} showed that a general electronic structure PES can be decomposed into local and body-ordered components by discretising Kohn–Sham DFT in a basis of localized orbitals, resulting in a tight-binding-like formulation. By equilibrating the new internal variable (i.e., achieving self-consistency), their local model is capable of representing non-trivial long-range effects.

A critical but often under-discussed challenge in modelling long-range interactions is the generation of suitable training data. 
As discussed in Section~\ref{better-data}, data generation is typically the most computationally expensive component of MLIP development, and this challenge is exacerbated for long-range physics. 
\textit{Ab initio} methods are inherently limited in accessible length and time scales, making it difficult to capture extended polarization or screening effects. 
Periodic boundary conditions further complicate the treatment of non-neutral systems, while dispersion interactions are often incorporated through semi-empirical corrections rather than directly learned from first-principles data. 
These limitations raise important questions about how best to construct training datasets that faithfully capture long-range physics, and whether current approaches can provide sufficiently transferable representations across diverse physical regimes.

From the above discussion two key questions emerge. First, can a single transferable approach capture long-range interactions across diverse physical scenarios, or will different forms of long-range physics require specialized models? Addressing this will require systematic benchmarking across a broad range of use cases, together with careful assessment of the trade-offs between accuracy, efficiency, and scaling with system size. Second, is there a unifying theoretical framework for long-range MLIPs? For short-range models, many seemingly distinct representations were shown to be understood within the ACE framework~\cite{drautz2019ace}. It is therefore natural to ask whether a similarly unified perspective could emerge for long-range interactions. While links between existing approaches have begun to appear~\cite{Kim2026,grgg2026}, such a unification has yet to be established.

\begin{figure}
    \centering
    \includegraphics[width=1\linewidth]{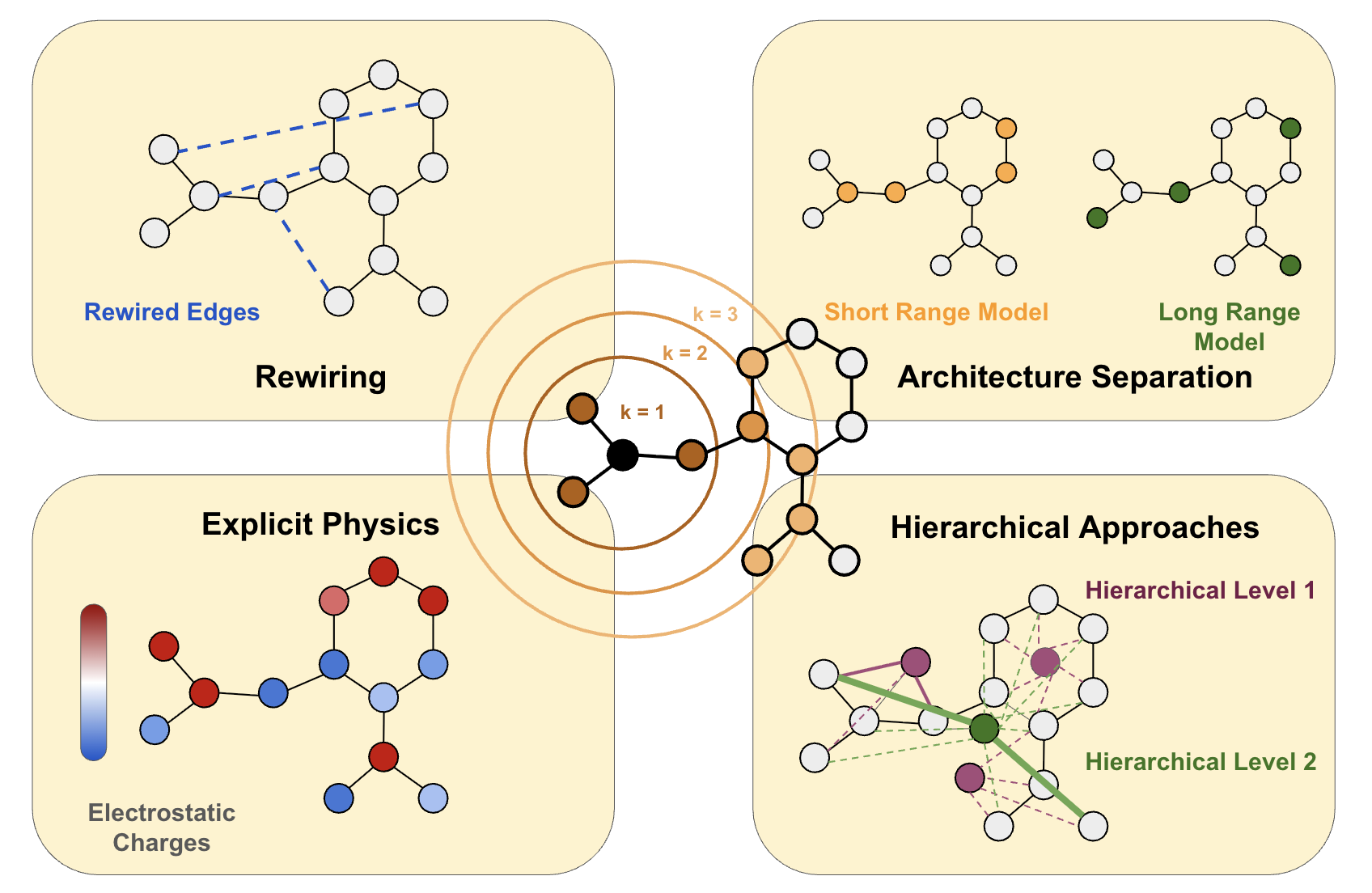}
    \caption{Overview of conceptual approaches to accurately capture long range interactions in GNNs while mitigating oversmoothing and oversquashing. Strategies include: Graph Rewiring (top left) for structural optimization; Dual Architectures (top right) for global-local separation; Hierarchical Approaches (bottom right) for multi-scale feature extraction; and incorporation of Physics-Informed Priors that encode the long-range governing physics (bottom left).}
    \label{fig:LR_overview}
\end{figure}


\section{Can MLIPs discover truly new physics?\label{new-physics}}

A defining aspiration of atomistic foundation models is that they should not merely interpolate within familiar chemistry but also generalise to new regimes, uncovering behaviours or mechanisms that are absent from their training data. Such capability can arise via two pathways. 

One pathway is generalisability: the ability to sustain reliability far outside the training distribution. 
Any striking generalisability present is sometimes referred to as 
emergent capability: qualitatively unexpected behaviour that is not expected from the training data. 
By analogy with LLMs, which seem able to do more than simply reproduce patterns in text, one may ask whether foundation MLIPs can likewise exhibit capabilities that go beyond straightforward interpolation or extrapolation.
Examples suggest that this may indeed be the case. The remarkable aspect of MACE-MP0~\cite{batatia_foundation_2024} was not only its benchmark accuracy, but that a model trained on near-equilibrium crystal structures could drive stable molecular dynamics simulations of liquids and interfaces. Similarly, models trained on small molecules and clusters that successfully transfer to bulk condensed-phase systems, such as esen-omol and MACE-OFF, exhibit this type of unexpected behaviour. Understanding when and why such an emergent capability arises is central to the promise of foundational MLIPs.

The other pillar is to be able to discover new underlying principles. This may take the form of interpretability, which is the capacity to turn learned representations into human-understandable physical insights. 
Or, this can be new phenomena or physics not encoded in the model or the training data.

Both pillars are tightly coupled to the inductive biases and expressivity of current architectures, as discussed in Sec.~\ref{definition}, \ref{models-or-data}, and \ref{long-range}. Local message-passing models excel at smooth interpolation, but often encode strong priors that limit exploration under extreme conditions or rare events. Scaled-up architectures promise broader coverage, yet, without appropriate physical constraints, may still fail to extrapolate meaningfully. Similarly, black-box expressivity helps match complex potential energy landscapes but makes it difficult to identify whether a model has discovered new physics or simply memorized subtle patterns in the data.

In this section, we examine the conditions under which MLIP frameworks might transcend interpolation and the challenges that currently hamper such behaviour. We also discuss early examples where data-driven models have begun and have the potential to illuminate physical phenomena beyond their training domain.

\subsection{What needs to happen for atomistic foundational models to discover new physics?}

MLIP frameworks today learn well within known chemistry and physics, but struggle outside training distributions, largely because they provide approximate representations of the underlying bonding physics in molecular systems by interpolating within the range of the training data~\cite{Anstine2023, liu_fine-tuning_2025}. This means that MLIP approaches are often poorly equipped to capture rare events or phenomena that occur under extreme conditions, such as high pressure, temperature, and strong fields~\cite{bartok_machine_2018}. Because such configurations are rare in typical datasets, standard mean-error objectives such as RMSE or MAE can become poor indicators of practical performance. These losses average over the dominant near-equilibrium configurations and may obscure catastrophic failures in rare but physically important regions of configuration space. Hence, models with low average errors may still fail to accurately reproduce key atomistic phenomena~\cite{botu_learning_2015}. Thus, there is a need to develop new error evaluation metrics, for example, force performance scores~\cite{liu_discrepancies_2023}. Much of the inability of models to generalise beyond training data is related to their expressivity and inductive biases, as discussed in Sec.~\ref{definition} and \ref{models-or-data}. In practice this means that benchmarks show good performance on thermodynamic physical observables, such as phonons~\cite{han_benchmarking_2025, pota_thermal_2025, loew2025} and prediction of formation energies, even for defect structures~\cite{jakob_universally_2025, berger_screening_2025}. This is, of course, conditioned on having good quality training or test data, as discussed in Sec.~\ref{better-data}, but the prediction of any other properties, that aren't dependent on energies or forces, is much more challenging~\cite{prakash_guided_2025}.

 A paradigmatic example of where limitations in training and inductive biases hamper extrapolation is calculating migration barriers, which is the basis to describe atomic diffusion performance in the material for batteries~\cite{lu_Hydrogenated_2024, roSeng_efficient_2016}, fuel cells~\cite{hirschfeld_first_2011}, and superionic conductors~\cite{yajima_correlated_2021}. The biased sampling for near-equilibrium in the training dataset causes potential energy surface softening where energy and force are underpredicted in complex atomic environments, especially in general high-energy states~\cite{deng_systematic_2025}. Consistent with these concerns, 
 benchmarks show that NEB errors of $0.74\pm0.06$~eV can arise even when the underlying system energy error is only $0.008$~eV --- two orders of magnitude smaller~\cite{maxson_MS2_2025}.
 A related example is the ability of MLIP approaches to accurately predict chemical reactivity, which is constrained when training data under-represent the target chemical space, particularly in attempts at automated reaction network exploration that introduces novel reactants and reagents across the periodic table~\cite{eckhoff_lifelong_2025}.



Current MLIP frameworks remain largely black-box models, with learned representations that are often difficult to map directly to established physical principles. For MLIPs to contribute meaningfully to the discovery of new physics, it is not sufficient that they make accurate predictions; their internal representations must also become accessible to physical interpretation, perhaps as parameters of an equation. Most state-of-the-art architectures, including artificial neural network (ANN)~\cite{zeng_deepmd-kit_2023}, GNN~\cite{batatia_mace_2022}, and emerging transformer-based approaches~\cite{zeni_mattergen_2024}, achieve high accuracy across diverse atomistic systems~\cite{kovacs_mace-off_2025}. However, this expressive flexibility often comes at the cost of interpretability, making it difficult to determine whether the models are learning physically meaningful relationships or simply complex statistical correlations~\cite{esders_analyzing_2025}.

Nevertheless, there is evidence that physically interpretable structure may arise in sufficiently large models. Recent work by Kreiman \textit{et al.}~\cite{kreiman2025transformersdiscovermolecularstructure}, for example, showed that a transformer model trained on the OMol25 dataset exhibited internal representations partially consistent with Coulombic interactions, despite no explicit encoding of Coulomb's law in the architecture. While such results remain preliminary, they suggest that large-scale models may implicitly recover aspects of underlying physical structure from data alone. Understanding when such behaviour emerges, and whether it can be systematically extracted and validated, remains an open challenge for the field.

\begin{figure}
    \centering
    \includegraphics[width=0.8\textwidth]{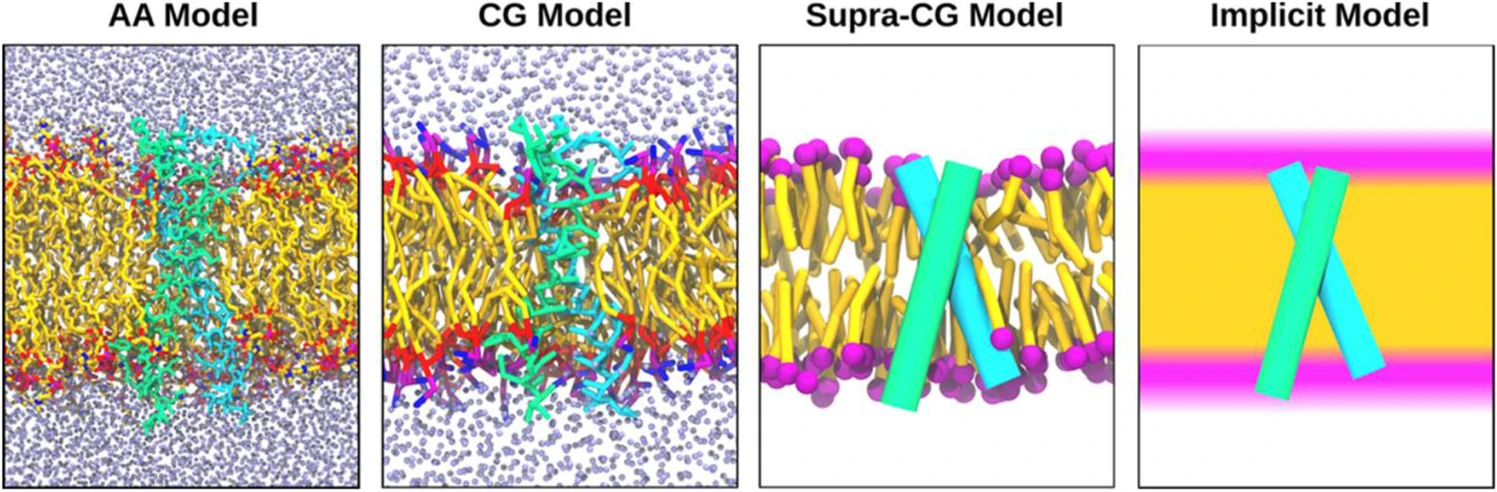}
    \caption{Different resolutions of lipid membranes. All-atom (AA) resolution explicitly considers all atoms. Coarse-grain (CG) resolution considers small atom groups and their associated hydrogens. Supra-CG resolution represents solvents implicitly and proteins and lipids as qualitative few-bead models. Implicit resolution further integrates out lipid molecules. Reprinted from~\cite{Carey2022}.}
    \label{figure:CoarseGraining}
\end{figure}

Looking from a more applied perspective, there are two major areas that would lead to new technological discoveries. Firstly, in the field of nanotechnology, the consensus has been shifting~\cite{Han2023, Stepanov2020, Choi2024, Muir2022, Biswas2023, Chan2022, Efferen2025, Barrier2024} towards the idea that more exotic device physics are to be found with long-range - strong correlation - effects included, together with detailed short-range descriptions. For example, a large tight-binding model, in combination with a far-stretching Coulomb interaction, describing a Moiré material.
Current ab-initio quantum chemistry approaches cannot reach the required system sizes at acceptable computational cost or accuracy, motivating the use of MLIP frameworks. Nevertheless, the required energy precision of these models of $\Delta E \sim 1$ meV is still not accurate enough for large system sizes or even heterostructures. Secondly, the area of drug research~\cite{Kobchikova2025} has, very recently, seen a sharp and impressive increase in MLIP framework applications. This is mainly, because coarse-grained techniques ~\cite{Joshi2020, Souza2021} and even all-atom ~\cite{Chaisson2023, Carey2022, Lee2018} approaches struggle with scaling and accuracy. As an example a cellular membrane model is shown in Fig.~\ref{figure:CoarseGraining} at these refinements - understanding this system is at the heart of drug mechanism research. Nevertheless, the main issue still remains - these systems are inherently disordered, meaning that quite often generalization, further from the training dataset, does not yield an accurate enough description. This is due to core molecular structures and charge localizations, as an example, among many others, not being present in the model.

\subsection{What new physics can be and are already being discovered?}

The issues listed above can be potentially addressed by recent algorithmic developments.

Uncertainty quantification can help via probabilistic frameworks~\cite{xie_advanced_2021, wen_uncertainty_2020}, ensemble-based methods~\cite{kurniawan_comparative_2025} and distance-based methods~\cite{janet_quantitative_2019, podryabinkin_accelerating_2019}. Such analyses can highlight where models fail, potentially pointing towards new physics. They can also facilitate on-the-fly learning~\cite{jacobsen2018fly}, where simulations understand when they are in a configuration beyond the training distribution and re-train the model on newly acquired ground-truth data for that configuration. We do note, however, that such on-the-fly schemes break the statistical ensemble and necessitate restarting of simulation to obtain accurate properties. Further advances have tried to tackle out-of-distribution data by augmenting the MLIP architecture with the inclusion of Hessians~\cite{rodriguez_does_2025}.

There have also been efforts to increase model expressivity - for example, a NequIP model has been combined with the direct prediction of polaronic sites and occupancies learned from VASP data~\cite{birschitzky_machine_2025}. There are also models which consider the magnetic moment of each atom~\cite{deng_chgnet_2023}, introduce a chemically informed force field contribution~\cite{ple_force-field-enhanced_2023}, directly learn pair-wise energy contributions~\cite{chun_learning_2025}, and introduce universal equations-of-state constraints~\cite{hu_global_2025}. 

Including multimodal training data has also been explored to improve the generalizability of models. Recent work ~\cite{gumber_going_2025} has looked at refining an initial MLIP trained on DFT/DFT+U data by improving its ability to replicate experimentally derived EXAFS spectra. Another approach used experimentally derived mechanical properties and lattice parameters to perform a fused data learning strategy~\cite{rocken_accurate_2024}. Both of these approaches have shown improvement in the models' predictive capabilities on the experimental labels used, as well as systematic improvement of all other properties tested.

Explainable AI (XAI)~\cite{lundberg_local_2020} enables the troubleshooting of model performance when predictions are poor, provides physical interpretation when models perform well, and could ultimately enable the discovery of new physics by bridging data-driven predictions with underlying physical principles~\cite{oviedo_interpretable_2022}. LIME~\cite{ribeiro_why_2016} and SHAP~\cite{kuhn_value_1953, lundberg_unified_2017} are widely used and have been successfully applied to ANN models, offering valuable insights into models used for predicting the dielectric constants of crystals~\cite{morita_modeling_2020} and for accelerating materials discovery~\cite{verissimo_integrating_2025, dangayach_machine_2025}. Efforts have also been made to extend these explainability frameworks to GNN approaches, where model interpretability remains an ongoing challenge. Recent developments, such as GNNShap~\cite{akkas_gnnshap_2024} and GraphSHAP-IQ~\cite{muschalik_exact_2025}, represent promising advances towards attributing GNN predictions, paving the way for more transparent and physically interpretable MLIP frameworks.

In practice, a couple of studies~\cite{Rowe2017, Thiemann2020, Siddiqui2024} have used MLIP frameworks for some of the most promising materials in nanotechnology - structures of graphene and transition metal dichalcogenide monolayers. They have been shown to be successful in predicting the vibrational properties - short-range characteristics -, at monolayer level, meanwhile providing a significant speed-up, in comparison to DFT, and, thus, opening up a pathway to study heterostructure system sizes. Additionally, charge density waves~\cite{Huang2017, Xi2015, Duvjir2018, Cheung2024} have been shown to be a long-range - correlation based - property of significant importance in these materials in order to analyse their various electronic phases, among these alternating superconducting and insulating ones. Very recently, they have been successfully studied~\cite{Rivano2025}, although in a size-limited fashion, via the help of an MLIP framework for the case of a monolayer and bilayer NbSe$_2$. The thermal properties of heterostructure alloys of doped MoS$_2$ and WS$_2$ monolayers have been studied~\cite{MarmolejoTejada2022} this way as well. This has primarily been achieved due to the developments in MLIP frameworks~\cite{Anstine2023} for better description of long-range contributions and scaling, whilst not being possible via standard ab-initio quantum chemistry approaches due to computational cost requirements. At this point, accurate simulations of Moiré and more complex heterostructure versions of such systems have yet to be conducted due to scaling issues, and other approaches are in early stages~\cite{Wang2024, Polshyn2021, Goodwin2022, Zhang2025} of development.

In addition, whilst there are a plethora of issues to be addressed before any MLIP approach would be able to accurately simulate biological systems relevant in drug development, there has been recent software development~\cite{Wang2024MD} of very large protein modelling (up to around $15000$ atoms), thus paving a potential pathway for characterizing large lipid layers as well. Hence, whilst research in practically vital areas is in early stages, great progress has already been made, and further work on scaling and precision would be quite certain to yield much more exotic, practically useful physics in these systems, whilst not being an unreasonable task.

Present-day MLIPs are beginning to demonstrate forms of extrapolative behaviour that would previously have been considered implausible, particularly in transferring across phases, scales, and chemical environments. However, most current successes still reflect sophisticated interpolation over broad datasets rather than autonomous discovery of fundamentally new physical laws. The more immediate scientific opportunity may therefore lie in using foundation MLIPs as hypothesis-generation and exploration tools: enabling access to regimes, system sizes, and collective phenomena that were previously computationally inaccessible. Whether such systems can ultimately progress from predictive models to genuine engines of scientific discovery will depend on advances in uncertainty quantification, interpretability, physically informed architectures, and multimodal integration with experiment.

\section{Can MLIPs scale to do more useful simulations?\label{scaling-mlips}}
In the context of MLIPs, scalability may refer to the ability to handle larger atomistic systems, longer simulation timescales, larger and more expressive models, or higher-throughput ensemble simulations.
Current foundation models, such as MACE-MP~\cite{batatia2025foundation}, have shown success in many applications, ranging from solid-state electrolytes~\cite{bertani_atomic_2025} and heterogeneous catalysis~\cite{equiformerv2} to organic drug-like molecules~\cite{kovacs_mace-off_2025}. However, large parameter counts needed to model the complexities of many chemical species lead to substantially longer energy and force evaluation times, compared to more lightweight and problem-specific MLIPs. 
We consider that if an MLIP is not at least an order of magnitude faster than the most efficient DFT alternative, then it is questionable if it is preferable to the \textit{ab initio} alternative.
Moreover, large foundation models often become memory-limited in large-scale simulations. Even on modern accelerators such as NVIDIA H200 GPUs, simulations containing more than $\sim$50k atoms may become unfeasible.
These limitations pose a critical barrier for applications involving long-timescale simulations, ensembles of trajectories, or systems containing thousands of atoms.
Numerous efforts to accelerate the scalability of MLIPs are currently being pursued in parallel, focusing on improving model architectures, optimizing performance for specific hardware, and enhancing treatment of MD time steps.
In this section, we review the approaches developed to address these challenges and outline potential directions for future improvements.

\subsection{Efficient architectures of foundation models} 

The motivation for building foundation machine-learning interatomic potentials (MLIPs) is to deploy a single large model that spans a broad chemical and structural domain, thereby avoiding the need to train a separate model for every system. They also tend to be more robust to atypical configurations—defects, high temperatures, and large strains, reducing the risk of pathological extrapolation. The trade-off is that foundation models typically have larger parameter counts and activation footprints, making per-step inference slower and more memory-intensive than lightweight, system-specific models. The computational cost of modern equivariant MLIPs is closely tied to how rotational symmetry is enforced mathematically. While linear ACE-type approaches remain attractive baselines~\cite{kovacs2021linear, witt2023acepotentials}, the top-performing entries on contemporary benchmarks~\cite{riebesell2025framework} are dominated by \(E(3)\)-equivariant graph neural networks (GNNs) such as eSEN-30M-OAM, Nequip-OAM-L, MACE-MPA-0and SevenNet-MF-ompa~\cite{fu_learning_2023,batzner20223, tan2025high,batatia_mace_2022, batatia2025foundation,Kim2024MultifidelityMLIP}. \(E(3)\)-equivariance enforces the exact transformation laws of energies, forces, and stresses, so the model ensure symmetry-correct outputs, often improving data efficiency and generalization.

Formally, let $X$ and $Y$ be representation spaces, and let $G$ be a symmetry group acting linearly on $X$ and $Y$ via representations
$\rho_X: G \to GL(X)$ and $\rho_Y: G \to GL(Y)$. We say that $f: X \to Y$ is \emph{$G$-equivariant} if, for all $g \in G$ and $x \in X$,
\begin{equation}
f\!\big(\rho_X(g)\,x\big)=\rho_Y(g)\,f(x).
\label{eq:G-equivariant}
\end{equation}
In atomistic modelling, \(G\) is typically the Euclidean group \(E(3)=\mathbb{R}^3\!\rtimes\!\mathrm{SO}(3)\) (optionally extended by inversion). Different models adopt different architectures to achieve \(E(3)\)-equivariance, most commonly by representing features as direct sums of \(\mathrm{SO}(3)\) irreducible representations and composing them via Clebsch–Gordan (CG) products. Edge directions \(\hat{\mathbf r}_{ij}\) are encoded with spherical harmonics \(Y^{\ell}_{m}\), which form an irreducible $\mathrm{SO}(3)$ basis, under a rotation $R$, they transform as
\begin{equation}
Y_m^{\ell}(R \hat{\mathbf{r}})=\sum_{m^{\prime}=-\ell}^{\ell} D_{m m^{\prime}}^{(\ell)}(R) Y_{m^{\prime}}^{\ell}(\hat{\mathbf{r}}),
\end{equation}
where $D^{(\ell)}(R)$ is the Wigner $D$-matrix (the $\ell$-th irreducible representation of $\mathrm{SO}(3)$). The Clebsch–Gordan coefficients define an $\mathrm{SO}(3)$-equivariant bilinear map from the tensor-product representation to irreducible representations:
\begin{equation}
\left(a^{\ell_1} \otimes b^{\ell_2}\right)_m^{\ell}=\sum_{m_1, m_2} C_{\ell_1 m_1, \ell_2 m_2}^{\ell m} a_{m_1}^{\ell_1} b_{m_2}^{\ell_2} .
\end{equation}
Consequently, CG-coupled tensor-product layers, followed by learnable mixing in multiplicity spaces, guarantee equivariance by construction.
This scheme is highly expressive — it captures fine angular structure and many-body couplings, but increases compute and memory: \(\ell>0\) blocks introduce \((2\ell+1)\) channels and dense CG contractions, with cost growing in \(\ell_{\max}\), neighbour count, and widths. However, recently proposed models such as So3krates~\cite{frank2023so3krates,fumc2024} or GotenNet~\cite{aykent2025gotennet} pursue \(E(3)\)-equivariance with compact, CG-free operators (e.g., inner-product–parameterized steerable maps), offering a lighter path that shows strong potential for accuracy–efficiency trade-offs. This lightweight direction appears promising for retaining accuracy while lowering computational and memory costs.

A different route to efficiency is to forgo equivariant tensor-product layers entirely. Non-equivariant or invariant-feature architectures, such as Orb~\cite{rhodes_orb-v3_2025}, PET~\cite{pc2023,mazitov_pet-mad_2025}, and AllScAIP~\cite{qwku2026}, operate on dense matrix operations by construction and can therefore directly benefit from heavily optimized standard attention kernels (e.g. FlashAttention~\cite{dferr2022}) and commodity hardware (dense generalized matrix multiplication on tensor cores) without requiring specialized equivariant kernel implementations. Such architectures trade physical priors for (theoretical) efficiency and instead learn symmetries from data~\cite{lpc2024}, although the practical impact of residual symmetry error is an open question.

\subsection{Distillation of MLIPs from foundation models}

One of the factors that enables modern MLIPs to achieve generalization across chemical domains is their size.
Common foundation models (FMs) are often built on complex graph-based architectures and have millions of parameters, making them $10^{3}$ to $10^{4}$ times more expensive than a classical force field ~\cite{jacobs_practical_2025}. 
The high computational cost makes inference challenging for researchers without access to HPC resources.
Knowledge distillation (KD), originally proposed by Hinton \textit{et al.}~\cite{hinton_distilling_2015}, offers a compelling solution to compress heavy models into more efficient ones. In the context of MLIPs, Amin \textit{et al.}~\cite{amin2025towards} showed that large-scale FMs could be distilled into smaller, specialized MLIPs (20-50$\times$ faster) through a distillation framework by matching the Hessians of the teacher and student model.
Additionally, Gardner \textit{et al.}~\cite{gardner_distillation_2025} show that distilling a briefly fine-tuned GNN FM into compact GNN or ACE student models delivers $\sim\!10{\times}$ and $\sim\!100{\times}$ faster inference, respectively, while keeping force MAE within $\sim\!5$–$15\%$ of the teacher. 
This pushed practical MD from the FM’s $\sim\!10^{3}$-atom ceiling to $10^{4}$–$10^{5}$ atoms with the GNN students, and up to $10^{6}$ atoms with ACE, without significant loss of fidelity.

In some studies, researchers have found that student models can surpass the performance of their teacher model~\cite{taniguchi_knowledge_2025}, illustrating KD as an effective regularization tool for MLIPs~\cite{matin_teacher-student_2025}. 
The teacher's knowledge, especially when averaged or complemented with additional signals like Hessians, can prevent the student from overfitting to the noise. 
For instance, Matin \textit{et al.}~\cite{matin_teacher-student_2025} observed up to 10\% lower force MAE in the student model when learned from an ensemble of teachers. 

Beyond runtime, KD also reduces the demand for new DFT labels.
A common workflow is to briefly fine-tune the teacher FM on only $\!10\sim\!10^{2}$ high-fidelity DFT labels and then generate a large synthetic set of structures that are labelled cheaply by the teacher. 
The student is trained on these soft targets (energies/forces), often yielding an order-of-magnitude ($\sim 10$–$100\times$) reduction in additional DFT calculations while retaining near-teacher accuracy~\cite{radova_fine-tuning_2025, gardner_distillation_2025}.


    

\subsection{Hardware-related speed-up}

Most modern MLIPs are based on NNs that map efficiently onto GPUs due to their high degree of parallelism and memory bandwidth. For large systems or long molecular dynamics (MD) trajectories, GPUs typically provide significant speedups compared to CPUs\cite{leimeroth_machine-learning_2025}. Currently, one of the most efficient frameworks for large-scale atomistic simulations is the C++-based Large-scale Atomic/Molecular Massively Parallel Simulator (LAMMPS)~\cite{thompson_lammps_2022}, which offers superior performance relative to Python-based MD codes and supports CPUs as well as NVIDIA, AMD, and Intel GPUs. Further acceleration can be achieved through the Kokkos package (LAMMPS–KOKKOS)~\cite{johansson_lammps-kokkos_2025}, which provides performance portability and minimizes CPU–GPU data transfer overhead.

Many recent MLIPs utilize equivariant features implemented through the e3nn framework~\cite{geiger2022e3nn}. Substantial speedups have been obtained with NVIDIA’s cuEquivariance package~\cite{cueq2024}, which introduces CUDA-specific kernels for e3nn-based models such as NequIP and MACE but remains restricted to NVIDIA hardware. The more recently released open-source OpenEquivariance library~\cite{bharadwaj_efficient_2025} provides equivalent functionality while supporting both NVIDIA and AMD GPUs, achieving up to 1.3$\times$ speedup over cuEquivariance and more than 10$\times$ over standard e3nn.

Nevertheless, CPU-based simulations can still remain competitive for systems of smaller sizes and/or massively parallel workloads, utilized for instance in gas–surface dynamics~\cite{stark_benchmarking_2024} or non-NN-based MLIPs (e.g., ACE~\cite{witt2023acepotentials}).

\subsection{Future directions}

\begin{figure}[h!]
    \centering
    \includegraphics[width=1.0\linewidth]{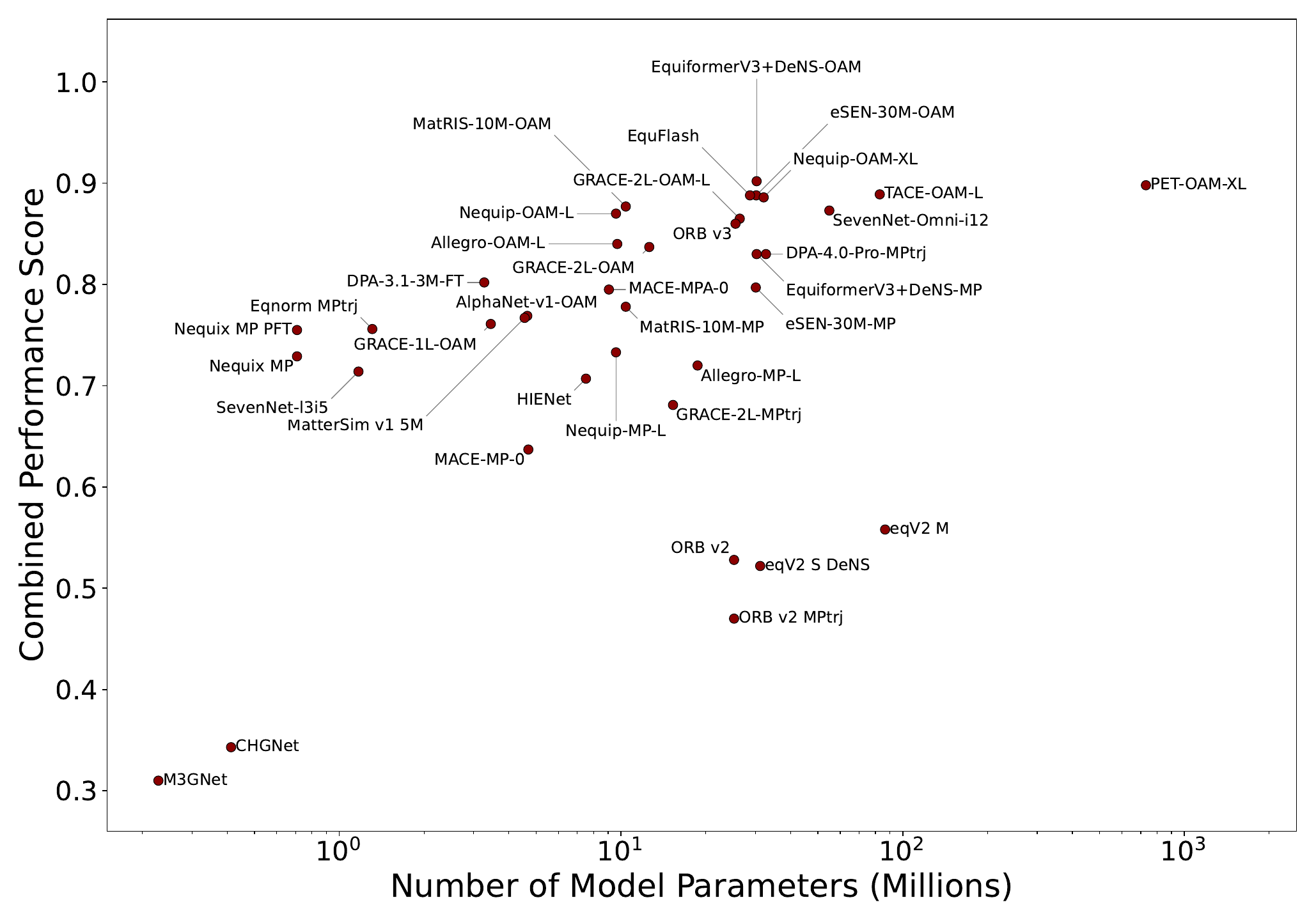}
    \caption{Combined performance score against the model size (number of model parameters) obtained with different foundation MLIPs, based on Matbench Discovery~\cite{riebesell2025framework} benchmark website (\url{https://matbench-discovery.materialsproject.org}, June 2026).}
    \label{fig:matbench_mlips}
\end{figure}

Large-scale atomistic simulations with MLIPs are already feasible, as demonstrated, for instance, by Musaelian \textit{et al.} in modelling large, biological systems~\cite{musaelian_scaling_2023}. However, such simulations still require substantial computational resources and often involve accuracy trade-offs. Developing accurate foundation MLIPs remains essential, but scalability and computational efficiency must not be overlooked, as they are also needed to enable large-scale simulations to study realistic systems within computing capabilities accessible to researchers worldwide.

As outlined in this section, to achieve more realistic and efficient models, future efforts should focus on improving inference efficiency through architectural innovation, faster dynamics schemes, and hardware-specific developments. Currently, the efficiency of MLIP models can often be compared only by model size or by other, architecture-related metrics (as shown in Fig.~\ref{fig:matbench_mlips}). However, it is not enough to reliably assess and compare the model efficiency.
Thus, we encourage more open reporting of model inference performance and the inclusion of efficiency metrics, especially in popular benchmarks such as Matbench Discovery~\cite{riebesell2025framework}.

Notably, the discussion above focuses almost entirely on inference-time efficiency. The complementary challenge of scaling training, for instance efficient batching of diverse structures~\cite{fpsc2025pre}, is equally critical. Model development requires training a large number of models, and can only keep pace with increasingly large datasets if training remains accessible and scalable.

Mixture-of-Experts (MoE) architectures offer another avenue for scaling model capacity without proportionally increasing inference cost. UMA~\cite{wdfgsbagklmscdrsuz2025} employs a Mixture-of-Linear-Experts (MoLE) approach that activates only a fraction of parameters per structure, and a recent systematic study~\cite{lzpezw2026} shows that sparse activation with shared experts and element-wise routing yields substantial accuracy gains on OMol25, OMat24, and OC20M benchmarks. More broadly, the MLIP community stands to benefit significantly from the fast-moving and well-funded LLM research ecosystem: FlashAttention, MoE routing, quantization, speculative decoding, and efficient serving infrastructure are all being developed at scale for language models and can, in many cases, be transferred to atomistic architectures with relatively modest adaptation. This is particularly true for MLIP architecture that trade domain-specific architectural features, such as equivariance, for simple standardized building blocks.

 Ultimately, the long-term utility of foundation MLIPs will depend not only on their predictive accuracy, but also on whether their computational cost scales favourably enough to enable scientifically relevant system sizes, timescales, and ensemble sampling. Achieving this balance will likely require co-design across architectures, hardware, training strategies, and simulation algorithms rather than progress along any single axis alone.

\section{How do we know if our MLIPs are any good? \label{benchmarks}}

\begin{figure}[h!]
    \centering
    \includegraphics[width=1.0\linewidth]{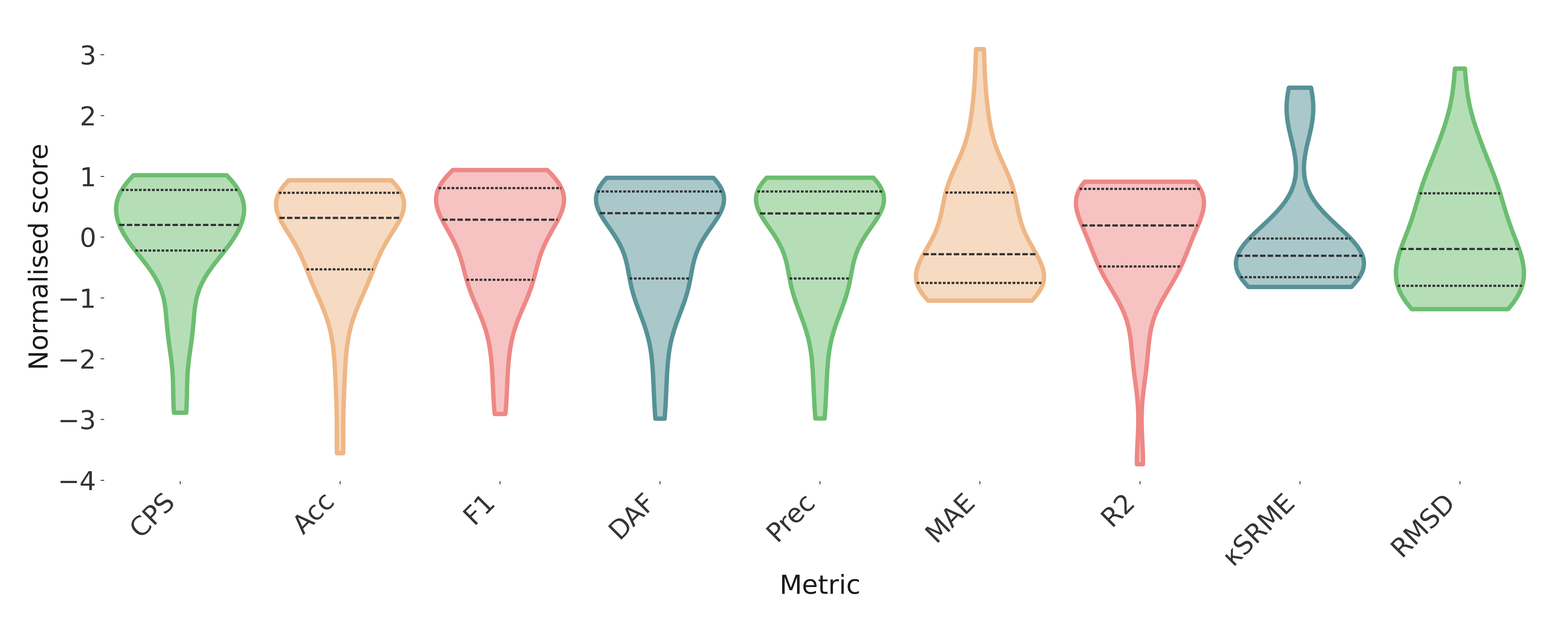}
    \caption{Distribution of the metric rankings from the leaderboard of the Matbench Discovery~\cite{riebesell2025framework} benchmark website (\url{https://matbench-discovery.materialsproject.org}, June 2026).}
    \label{fig:matbench_mlips_leader}
\end{figure}

An overarching question covering all of the subjects we have covered is: how do we know if a particular model can do what we want? We may want to do a large simulation of a well-known system, or explore some new exotic physics; perhaps our system is highly charged --- how can we decide which of the zoo of MLIPs is best for the task? Which MLIPs are most scalable, which include the right inductive biases for our task, can they be fine-tuned for our purposes?  This is where the topic of benchmarking becomes very important.

To frame the discussion of benchmarks, we adopt the terminology of Jablonka and co-workers~\cite{alampara2025lessons}, placing benchmarks on a scale from representational to pragmatic. At the representational end of the spectrum are benchmarks that probe intrinsic or architecture-level properties of a model, such as parameter count, scaling behaviour, or inference cost. Pragmatic benchmarks, on the other hand, are deliberately constructed to mimic practical situations and aid decision making; for example, how long a stable MD simulation at elevated temperature runs before the energy diverges. Both ends of the spectrum have attractive qualities, and both have drawbacks that should be acknowledged.

Pragmatic benchmarks are attractive because they directly probe properties of direct interest, providing rankings that aid model selection. However, pragmatic benchmarks often directly influence the systems that they measure; a well-known example is how the h-index measure, which can affect the choices of academic researchers. More generally, this is formalized as Goodhart's Law: ``Any observed statistical regularity will tend to collapse once pressure is placed upon it for control purposes''~\cite{goodhart1975problems}.

Representational benchmarks, on the other hand, are less susceptible to hacking than pragmatic benchmarks. However, it can be difficult to identify the inherent properties of the model that can be measured to assess practical utility. Additionally, certain properties are easier to measure than others, and this can lead to the McNamara fallacy. ``But when the McNamara discipline is applied too literally, the first step is to measure whatever can be easily measured. The second step is to disregard that which can't easily be measured or given a quantitative value. The third step is to presume that what can't be measured easily really isn't important. The fourth step is to say that what can't be easily measured really doesn't exist. This is suicide.''~\cite{mcnamarafallacy}

The discipline of benchmarking MLIPs is new and evolving; certain benchmarks 
have nonetheless become \textit{de facto} standards, while new ones continue 
to emerge to address perceived limitations. The dominant benchmark is currently 
Matbench Discovery~\cite{riebesell2025framework}, upon which we have drawn
extensively throughout this article. It blends representational and pragmatic 
measurements and provides a single unified metric composed from all constituent 
scores. In this sense, Matbench Discovery has become the MNIST~\cite{deng2012mnist} 
of MLIPs: widely adopted, practically useful, and increasingly a victim of its 
own success. As with MNIST, sustained community pressure on a single metric 
drives models to aggregate near peak performance, at which point incremental 
gains may reflect inconsistencies or uncertainty in the reference data rather 
than genuine model improvement. This is consistent with what Figure~\ref{fig:matbench_mlips_leader} 
reveals: at the top of the leaderboard, the distributions of performance metrics are becoming 
tightly clustered.
Importantly, the Matbench Discovery website does provide useful tools to interrogate models along different dimensions interactively, which means that researchers can adapt the frameworks for their own purposes.

Newer benchmarks have started to appear for MLIP evaluation. MLIP-Arena is designed to test the physical fidelity of models~\cite{chiang2025mlip}; ML-PEG prioritises a diverse set of simulation tasks and system types \cite{Kasoar_ml-peg}; the matPES benchmark deliberately restricts comparison to models trained on a specific dataset\cite{kaplan2025foundational}. The OMol25 dataset has also included additional physics-based evaluations~\cite{huggingfaceFAIRChemistry}. Probing how well the inductive biases and training data of the model allow it to capture properties such as diatomic dissociation curves, reactivity, and performance in extreme conditions. These criteria may be closely associated with many of the open questions that we outlined, such as a model's ability to capture new physics (Sec.~\ref{new-physics}), scale (Sec.~\ref{scaling-mlips}), and treat long-range interactions (Sec.~\ref{long-range}).
Other efforts have also been made to probe the generalization power of MLIP models using structure-based metrics to construct test tasks that are purposely beyond the distribution of the training data~\cite{omee2024structure}. Furthermore, the generation of new datasets, for example those with charged cells, enables the benchmarking of emerging MLIP features \cite{ko2025fast}.

It seems clear that benchmarking will remain an active, critical, and often controversial topic in the development of any type of computer model. While there are already several excellent frameworks available, the subject and the requirements of users will develop; so too will the benchmarks. New tools for capturing the environmental impact of model training and inference also make possible testing for more responsible application~\cite{lottick2019energy, walker2026carbon}.  We believe that the fundamental properties of a foundation model should be considered when designing benchmarks to assess and compare models. Furthermore, we echo the calls from Alampara \textit{et al.}~for maximum transparency and documentation of limitations of benchmarks to avoid the traps of Goodhart's Law or the McNamara fallacy~\cite{alampara2025lessons}. 

\section*{Conclusion}

This is an exciting time to be involved in molecular simulation. The kinds of experiments which only a decade ago seemed to be infeasibly out of reach; modelling complex multi-component systems, featuring rare-events, with little to no existing training data; now seem to be plausible with the emergence of foundation model MLIPs. This further opens up the promise of being able to conduct \textit{bone fide}, bottom-up atomisitic design of new chemistry and materials.

However, the rapid emergence and rise of MLIPs means that there are many open questions and avenues to explore. In this article we have covered a set of what we see as the most important questions. As evidenced in our examination of each of these questions there are multiple potential routes that can be explored. By examining and articulating the diverse options we aim to promote a critical engagement in the research community and to avoid the trap of easy acceptance of orthodoxy. We believe that with intense and open examination of these open questions, the full potential of ML for atomistic molecular simulation will be realised.

\section*{Acknowledgments}

We thank Shyue Ping Ong and G\'{a}bor Cs\'{a}nyi for their constructive feedback on the manuscript. KTB acknowledges funding from EPSRC (EP/Y014405/1 and EP/Y000552/1). CB and MAHW are funded by a UCL start-up package. K.W. acknowledges funding from EPSRC (EP/R513143/1 and EP/W524335/1). MT is funded by an Ada Lovelace Centre PhD studentship. AIchemy (AMG, WGS, KTB) acknowledges the funding support by the EPSRC Grants EP/Y028775/1 and EP/Y028759/1. IV is funded by a Royal Society doctoral studentship from grant URF-R1-191292. I.C. is supported by EPSRC (EP/Y020790/1). T.R. is a Royal Society funded PhD student. A.Y.I acknowledges funding from the Daphne Jackson Trust and UKRI. MFL acknowledges funding from the German Research Foundation (DFG) under project number 544947822.
\bibliographystyle{iopart-num}  
\bibliography{bibliography_unified,ml}

\end{document}